\documentclass[aps,prb,twocolumn,fleqn,eqsecnum,floatfix]{revtex4}

\usepackage{amssymb}
\usepackage{amsmath}
\usepackage{graphicx}

\begin{document}

\author{Adolfo Avella}
\email[E-mail: ]{avella@sa.infn.it} \homepage[Group Homepage:
]{http://scs.sa.infn.it}
\author{Ferdinando Mancini}

\affiliation{Dipartimento di Fisica ``E.R. Caianiello'' -
Laboratorio Regionale SuperMat, INFM \\ Universit\`{a} degli Studi
di Salerno, I-84081 Baronissi (SA), Italy}

\title{Exact solution of the one-dimensional spin-$\frac32$ Ising model in magnetic field}

\begin{abstract}
In this paper, we study the Ising model with general spin $S$ in
presence of an external magnetic field by means of the equations of
motion method and of the Green's function formalism. First, the
model is shown to be isomorphic to a fermionic one constituted of
$2S$ species of localized particles interacting via an intersite
Coulomb interaction. Then, an exact solution is found, for any
dimension, in terms of a finite, complete set of eigenoperators of
the latter Hamiltonian and of the corresponding eigenenergies. This
explicit knowledge makes possible writing exact expressions for the
corresponding Green's function and correlation functions, which turn
out to depend on a finite set of parameters to be self-consistently
determined. Finally, we present an original procedure, based on
algebraic constraints, to exactly fix these latter parameters in the
case of dimension $1$ and spin $\frac32$. For this latter case and,
just for comparison, for the cases of dimension $1$ and spin
$\frac12$ [F. Mancini, Eur. Phys. J. B \textbf{45}, 497 (2005)] and
spin $1$ [F. Mancini, Eur. Phys. J. B \textbf{47}, 527 (2005)],
relevant properties such as magnetization $\langle S \rangle$ and
square magnetic moment $\langle S^2 \rangle$, susceptibility and
specific heat are reported as functions of temperature and external
magnetic field both for ferromagnetic and antiferromagnetic
couplings. It is worth noticing the use we made of composite
operators describing occupation transitions among the $3$ species of
localized particles and the related study of single, double and
triple occupancy per site.
\end{abstract}

\date{\today}

\maketitle

\section{Introduction}

In a previous article \cite{Mancini_05}, we have shown that a system
constituted of $q$ species of particles, satisfying Fermi
statistics, subject to finite-range interactions and localized on
the sites of a Bravais lattice, is exactly solvable in any
dimension. By \emph{exactly solvable} we mean that it is always
possible finding a complete, finite set of eigenenergies and
eigenoperators of the Hamiltonian which closes the hierarchy of the
equations of motion. Then, formal exact expressions for the
corresponding Green's functions and correlation functions can be
derived. The solution is only formal as these latter turn out to
depend on a finite set of parameters to be self-consistently
determined. In Refs.~\onlinecite{Mancini_05b,Mancini_05a}, we have
shown how is possible to exactly fix such parameters, by means of
algebraic constraints, for the one-dimensional $q=1$ and $q=2$
cases, respectively.

In Ref.~\onlinecite{Mancini_05}, we have also shown that this system
is isomorphic to a spin-$\frac{q}2$ Ising-like model in presence of
an external magnetic field. According to this, the exact knowledge
of a complete set of eigenoperators and eigenenergies of the system
in any dimension acquires an evident relevance with respect to the
hoary problem of solving the two-dimensional (in presence of an
external magnetic field) and the three-dimensional Ising models. In
particular, the exact knowledge of the eigenenergies of the system
can shed some light on the energy scales ruling the physical
properties and the magnetic response of the system and can find an
application as unbiased check for the approximate solutions present
in the literature. Moreover, this approach can eventually open a new
route in the quest for an exact solution for these systems in higher
dimensions as we have shown that it is always possible finding a
formal exact expressions for the corresponding Green's functions and
correlation functions. The solution is only formal as these latter
turn out to depend on a finite set of parameters to be
self-consistently determined. We have also shown how is possible to
exactly fix such parameters, by means of algebraic constraints, for
the one-dimensional case and we are now working on the possibility
to use other algebraic constraints and topological relations for
higher dimensions.

In this article, we apply this formulation to the one-dimensional
$q=3$ case. This model is isomorphic to the one-dimensional Ising
model for spin $S=\frac32$ in presence of an external magnetic
field. This latter model can be exactly solved by means of the
transfer matrix method which reduces the solution to an eigenvalue
problem of the fourth order (generally, we have an eigenvalue
problem of the $2S+1$-th order). Unfortunately, the analytic
solutions of a fourth order equation are well known, but absolutely
untractable and any result, along this way, could be obtained only
by means of numerical techniques, which surely do not facilitate a
microscopic understanding of the properties of the system (see, for
instance, Ref.~\onlinecite{Aydmer_05}). In 1967, by means of a
perturbation method, Suzuki, Tsujiyama and Katsura managed to reduce
the order of the problem to the largest integer smaller than $S+1$
in the case of zero external magnetic field and exactly computed the
energy, the specific heat and susceptibility \cite{Suzuki_67}. On
year later, Obokata and Oguchi \cite{Obokata_68} managed to apply
the Bethe approximation, which becomes exact in one dimension,  to
the system and to recover the same exact results obtained by Suzuki.
These latter were also recovered, few years later, by Silver and
Frankel \cite{Silver_71}, who managed to formulate the problem as a
difference equation of order $2S+1$ by an inductive technique and to
reduce the order to the largest integer smaller than $S+1$ in the
case of zero external magnetic field.

The spin-$\frac32$ Ising model in one dimension shows magnetic
plateaus \cite{Chen_03,Ohanyan_03,Aydmer_04,Aydmer_05a}, i.e.
topological quantization of the magnetization at the ground state of
the system due to magnetic excitations, leading to the qualitatively
same structures of the magnetization profiles of the Heisenberg
model. According to this, the study of this classical system can
shed some light on the plateau mechanism and if it has purely
quantum origin or can also depend on dimerization, frustration,
single-ion anisotropy or periodic field. These plateaus have been
predicted not only in theoretical study but also have been observed
in experimental studies. For example, Narumi et al.
\cite{Narumi_98,Narumi_98a} observed the magnetic plateaus in the
magnetization curve for both
$[Ni_2(Medpt)_2(\mu-ox)(H_2O)_2](ClO_4)_2 2H_2O$ and
$[Ni_2(Medpt)_2(\mu-ox)(\mu-N_3)](ClO_4)_{0.5} H_2O$. Goto et al.
\cite{Goto_01} reported the existence of the magnetization plateau
at $0.25$ in spin-1 3,3',5,5'-tetrakis (N-tert-butylaminxyl)
biphenyl (BIP-TENO). In three dimensions and with additional terms
in the Hamiltonian, the spin-$\frac32$ Ising model have been
initially introduced to give a qualitative description of phase
transition observed in the compound $DyVO_4$ \cite{Sivardiere_72}
and also to describe ternary mixtures \cite{Krinsky_75}. In
particular, the phase diagram is not well known in contrast to the
case $S=1$, which represents a special case of the Blume-Capel
\cite{Blume_66,Capel_66,Capel_67,Capel_67a} and of the
Blume-Emery-Griffiths \cite{Blume_71} models often used to study a
variety of interesting physical systems and, in particular,
$\mbox{}^3He-\mbox{}^4He$ mixtures, fluid mixtures and critical
phenomena. The mean field treatment \cite{Plascak_93} predicts that
the phase diagram differs for integer and half-odd-integer spins and
does not present any multicritical point. While
renormalization-group calculations
\cite{deOliveira_92,deOliveira_95} and Monte Carlo simulations
\cite{deSaBarreto_91} suggest the existence of a multicritical
point, transfer matrix and conformal invariance studies
\cite{Xavier_98} show that there is no multicritical point in the
phase diagram. Unfortunately, Bethe-Peierls
\cite{Kaneyoshi_92,LeGal_93} and two-spin-cluster approximations
\cite{Jurcisin_96} have not been performed in the low-temperature
region. In qualitative agreement with the mean-field analysis, we
can also find results from the self-consistent Ornstein-Zernike
approximation \cite{Grollau_02}.

In this manuscript, we present the exact solution of the model in
presence of an external magnetic field. In the first section, we
present the model. In the following section, we give the general
solution in terms of the eigenoperators and of the eigenenergies. In
the third section, we specialize the solution to the one-dimensional
$q=3$-$S=\frac32$ case. In the fourth section, we show how to close
the self-consistent equations and compute all relevant correlation
functions. In the fifth section, we present relevant properties such
as magnetization $\langle S \rangle$ and square magnetic moment
$\langle S^2 \rangle$, susceptibility and specific heat as functions
of the temperature and the external magnetic field both for
ferromagnetic and antiferromagnetic couplings. Concluding remarks
follow.

\section{The model}

We have analyzed a system constituted of $q$ species of interacting
particles obeying Fermi statistics, localized on the sites of a
Bravais lattice and whose dynamics is ruled by the following
grand-canonical Hamiltonian
\begin{equation} \label{2.1}
H = -\mu \sum_{\mathbf{i},a} n_{a}(i)+
\frac{1}{2}\sum_{\mathbf{ij},ab}V_{ab}(\mathbf{i,j})n_{a}(i)n_{b}(j)
\end{equation}
where $\mathbf{i}$ stands for the lattice vector $\mathbf{R}_{i}$
and $i=(\mathbf{i},t)$. $n_{a}(i)=c_{a}^{\dagger }(i)c_{a}(i)$ is
the particle density operator of particles of species $a$ at the
site $\mathbf{i}$. $c_{a}(i)$ and $c_{a}^{\dagger }(i)$ are
annihilation and creation operators, respectively, of particles of
species $a$ at the site $\mathbf{i}$. They satisfy canonical
anti-commutation relations
\begin{equation}\label{2.2}
\begin{split}
&\{c_{a}(\mathbf{i},t),c_{b}^\dagger(\mathbf{j},t)\}=\delta_{ab}\delta
_{\mathbf{ij}} \\
&\{c_{a}(\mathbf{i},t),c_{b}(\mathbf{j},t)\} =
\{c_{a}^\dagger(\mathbf{i},t),c_{b}^\dagger(\mathbf{j},t)\}=0
\end{split}
\end{equation}
$\mu$ is the chemical potential and $V_{ab}(\mathbf{i,j})$ is the
strength of the interaction between particles of species $a$ and $b$
at distance $|\mathbf{i}-\mathbf{j}|$. We have supposed that the
particles are frozen on the lattice sites as their masses are very
large and/or the interactions are so strong that the kinetic energy
is negligible.

In this manuscript, we have focused on the case in which the
interaction $V_{ab}(\mathbf{i,j})$ does not depend on the particle
species and is effective only between nearest-neighbor sites. That
is, $V_{ab}(\mathbf{i,j})=2dV\alpha _{\mathbf{ij}}$ where $d$ is the
dimensionality of the system, $\alpha _{\mathbf{ij}}$ is the
projector on the nearest-neighbor sites and $V$ is the bare strength
of the interaction. For a hyper-cubic lattice of lattice constant
$a$ we have
\begin{equation} \label{2.6}
\begin{split}
&\alpha _{\mathbf{ij}}=\frac{1}{N}\sum _{\mathbf{k}}e^{i\mathbf{k}
\cdot (\mathbf{R}_{i}-\mathbf{R}_{j})} \alpha ( \mathbf{k}) \\
&\alpha ( \mathbf{k})=\frac{1}{d}\sum _{n=1}^d \cos (k_{n}a)
\end{split}
\end{equation}
where $N$ is the number of sites.

Hereafter, for a generic operator $\Phi(i)$, we will use the
following notation
\begin{equation} \label{2.8}
\Phi^{\alpha }(\mathbf{i},t) = \sum_{\mathbf{j}}
\alpha_{\mathbf{ij}} \Phi(\mathbf{j},t)
\end{equation}. It is also useful to introduce the vectorial notation
\begin{equation}\label{2.3}
\begin{split}
&c(i)=\left(
\begin{array}{c}
c_{1}(i) \\
c_{2}(i) \\
\vdots \\
c_{q}(i)
\end{array}
\right)\\
&c^{\dagger }(i)=\left(
\begin{array}{cccc}
c_{1}^{\dagger }(i) & c_{2}^{\dagger}(i) & \cdots & c_{q}^{\dagger
}(i)
\end{array}
\right)
\end{split}
\end{equation}
and to define the total particle density
\begin{equation}\label{2.5}
n(i)=\sum_{a}n_{a}(i)=\sum_{a}c_{a}^{\dagger }(i)c_{a}(i)=c^{\dagger
}(i)c(i)
\end{equation}
Then, the Hamiltonian (\ref{2.1}) becomes
\begin{equation} \label{2.7}
H = -\mu \sum_{\mathbf{i}} n(i)+ d V \sum_{\mathbf{i}} n(i)
n^{\alpha}(i)
\end{equation}

Let us consider now the transformation
\begin{equation} \label{2.9}
n(i) = {\frac{q}{2}} + S(i)
\end{equation}
It is clear that
\begin{equation} \label{2.10}
\begin{split}
& n(i)=0 \Leftrightarrow S(i)=-q/2 \\
& n(i)=1 \Leftrightarrow S(i)=1-q/2 \\
\vdots \\
& n(i)=q-1 \Leftrightarrow S(i)=q/2-1 \\
& n(i)=q \Leftrightarrow S(i)=q/2
\end{split}
\end{equation}
Under the transformation (\ref{2.9}), the Hamiltonian (\ref{2.7})
can be cast in the form
\begin{equation}
H=-dJ\sum_{i}S(i)S^{\alpha }(i)-h\sum_{i}S(i)+E_{0} \label{2.11}
\end{equation}
where we defined
\begin{equation} \label{2.12}
\begin{split}
& J = -dV \\
& h = \mu -qdV \\
& E_{0} = {\frac{q}{2}}(-\mu +{\frac{q}{2}}dV)N
\end{split}
\end{equation}
The Hamiltonian (\ref{2.11}) is just the $d$-dimensional
spin-$\frac{q}{2}$ Ising model with nearest-neighbor interaction in
presence of an external magnetic field. In this manuscript, we have
chosen to use the particle notation (\ref{2.7}), but the results
that we have obtained are obviously valid for both the particle
(\ref{2.7}) and spin (\ref{2.11}) systems after the transformation
(\ref{2.9}) and the definitions (\ref{2.12}).

\section{Composite operators and equations of motion}

In order to apply the equations of motion method and the Green's
function formalism, we need to identify a suitable operatorial basis
\cite{Mancini_00,Mancini_04}. It is immediate to verify that the
particle density operator $n(i)$ has no time dependence
\begin{equation} \label{3.1}
\mathrm{i}{\frac{\partial }{\partial t}}n(i)=[n(i),H]=0
\end{equation}
According to this, the operator $n(i)$, although it would have been
a natural choice as component of the operatorial basis for such a
system, is not suitable for this purpose. Let us introduce, instead,
the following series of composite field operators
\begin{align} \label{3.1a}
&\psi_{p}(i)=c(i)[n^{\alpha }(i)]^{p-1} &&& p=1,2,\ldots
\end{align}
whose first element is just $c(i)$. These fields satisfy the
following hierarchy of equations of motion
\begin{equation} \label{3.2}
\mathrm{i}{\frac{\partial }{\partial t}} \psi _{p}(i) = -\mu
\psi_{p}(i) + 2dV \psi_{p+1}(i)
\end{equation}
Now, because of the anti-commutation relations (\ref{2.2}), it can
be shown that the operators $[n^{\alpha}(i)]^{p}$ satisfy the
following relation
\begin{equation} \label{3.3}
[n^{\alpha }(i)]^{p}=\sum_{m=1}^{2qd}A_{m}^{(p)}[n^{\alpha }(i)]^{m}
\end{equation}
where the coefficients $A_{m}^{(p)}$ are rational numbers that can
be easily determined after the algebra and the actual structure of
the lattice (see Appendix A for $d=1$ and $q=3$). Then, for
$p=2qd+1$, the hierarchy of equations of motion (\ref{3.2}) closes
as the additionally generated operator
$\psi_{2qd+2}(i)=c(i)[n^{\alpha }(i)]^{2qd+1}$ can be rewritten in
terms of the first $2qd+1$ elements of (\ref{3.1a}) through the
relation (\ref{3.3}).

According to this, the $n$-component composite field operator $\psi
(i)$, defined as
\begin{equation} \label{3.5}
\psi (i)=\left(
\begin{array}{c}
{\psi _{1}(i)} \\
{\psi _{2}(i)} \\
\vdots \\
{\psi _{n}(i)}
\end{array}
\right) =\left(
\begin{array}{c}
{c(i)} \\
{c(i)n^{\alpha }(i)} \\
\vdots \\
{c(i)[n^{\alpha }(i)]^{n-1}}
\end{array}
\right)
\end{equation}
where $n=2qd+1$, is an eigenoperator of the Hamiltonian (\ref{2.7})
as it satisfies the equation of motion
\begin{equation} \label{3.6}
\mathrm{i}{\frac{\partial }{{\partial t}}}\psi (i)=[\psi
(i),H]=\varepsilon \psi (i)
\end{equation}
where the $n\times n$ energy matrix $\varepsilon$ has the following
expression
\begin{equation} \label{3.7}
\varepsilon =\left(
\begin{array}{ccccccc}
-\mu & 2dV & 0 & \cdots & 0 \\
0 & -\mu & 2dV & \cdots & 0 \\
\vdots & \vdots & \vdots & \vdots & \vdots \\
0 & 0 & 0 & \cdots & 2dV \\
0 & 2dVA_{1}^{(n)} & 2dVA_{2}^{(n)} & \cdots & -\mu
+2dVA_{n-1}^{(n)}
\end{array}
\right)
\end{equation}
The eigenvalues $E_{n}$ of the energy matrix $\varepsilon$ read as
\begin{align} \label{3.8}
& E_{m}=-\mu +(m-1)V &&& m=1,2,\ldots,n
\end{align}
After (\ref{3.6}), we can claim to have formally, but exactly,
solved both Hamiltonians (\ref{2.7}) and (\ref{2.11}) as we have
determined for them a complete set of eigenoperators and eigenvalues
in any dimension $d$. The solution is only formal as we have still
to compute all correlation functions.

Let us define the thermal retarded Green's function
\begin{equation} \label{3.9}
G^{R}(i,j)= \langle R [\psi (i) \psi ^{\dagger }(j)] \rangle =
\theta (t_{i}-t_{j}) \langle \{\psi (i),\psi ^{\dagger }(j)\}\rangle
\end{equation}
where $\langle \cdots \rangle$ denotes the quantum-statistical
average over the grand canonical ensemble. By introducing the
Fourier transform
\begin{multline} \label{3.10}
G^{R}(i,j)=\\
=\frac{\mathrm{i}}{(2\pi )^{2d+1}}\iint d\omega d\mathbf{k} \,
e^{i\mathbf{k}\cdot (\mathbf{R} _{i}-\mathbf{R}_{j})-i\omega
(t_{i}-t_{j})}G^{R}(\mathbf{k},\omega )
\end{multline}
and by means of the equation (\ref{3.6}), we obtain the equation
\begin{equation}
[\omega -\varepsilon ]G^{R}(\mathbf{k},\omega )=I(\mathbf{k})
\label{3.11}
\end{equation}
where $I(\mathbf{k})$ is the Fourier transform of the normalization
matrix
\begin{equation} \label{3.12}
I(\mathbf{i,j})= \langle \{\psi (\mathbf{i},t),\psi ^{\dagger
}(\mathbf{j},t)\} \rangle
\end{equation}
Equation~(\ref{3.11}) gives
\begin{equation} \label{3.13}
G^{R}(\mathbf{k},\omega )=\sum_{m=1}^{n}\frac{\sigma
^{(m)}(\mathbf{k})}{\omega -E_{m}+\mathrm{i}{\delta }}
\end{equation}
The spectral density matrices $\sigma ^{(m)}(\mathbf{k})$ can be
calculated by means of the formula \cite{Mancini_00}
\begin{equation}  \label{3.14}
\sigma _{ab}^{(m)}(\mathbf{k})=\Omega _{am}\sum_{c}\Omega
_{mc}^{-1}I_{cb}(\mathbf{k})
\end{equation}
where $\Omega $ is the $n\times n$ matrix whose columns are the
eigenvectors of the matrix $\varepsilon$. The correlation functions
\begin{equation}  \label{3.17}
C(i,j)=\langle\psi (i)\psi ^{\dagger }(j)\rangle
\end{equation}
can be immediately calculated after (\ref{3.13}) and read as
\begin{equation}  \label{3.18}
C(\mathbf{k},\omega )=\pi \sum_{m=1}^{n}T_{m}\sigma ^{(m)}(\mathbf{k
})\delta (\omega -E_{m})
\end{equation}
where
\begin{equation}  \label{3.19}
T_{m}=1+\tanh \left( \frac{E_{m}}{2T} \right)
\end{equation}
$T$ is the temperature. By similar techniques we can calculate
multi-point correlation functions as
$C(i,j;l_{1},l_{2},\ldots,l_{s})=\langle \psi (i)\psi ^{\dagger
}(j)n(l_{1})n(l_{2})\cdots n(l_{s})\rangle $. For an example, see
Ref.~\onlinecite{Mancini_05b}.

Formally, (\ref{3.13}) and (\ref{3.18}) constitute the exact
solution of the problem under analysis. The solution is only formal
as the complete knowledge of the Green's function and of the
correlation functions is not fully achieved owing to unknown static
correlation functions appearing in the normalization matrix
$I(\mathbf{k})$ because of the complex non-canonical algebra
satisfied by the field $\psi(i)$. The unknown correlators are
expectation values of operators not belonging to the chosen basis
$\psi(i)$ and should be self-consistently calculated in order to
definitively conclude the analysis and get the complete exact
solution. Unfortunately, the derivation of a set of self-consistent
equations capable to determine the unknown correlators is not an
easy task at all. Moreover, it depends on the particular model under
analysis and, in particular, on its dimension $d$. In
Ref.~\onlinecite{Mancini_05}, we have developed a technique to
obtain a complete exact solution in the one-dimensional case ($d=1$)
by determining the unknown correlators by means of algebraic
constraints. Within this framework, we have studied the
$2$-state\cite{Mancini_05b} ($q=1$) and the
$3$-state\cite{Mancini_05a} ($q=2$) models. In this article, we
consider the one-dimensional $4$-state ($q=3$) model.

\section{The one-dimensional $4$-state model}

We now apply the general formulation given in the previous Section
to an infinite homogeneous chain ($d=1$) with $q=3$. The derivation
of the relation (\ref{3.3}) for this specific case and the explicit
expressions of the coefficients $A_{m}^{(p)}$ are given in Appendix
A. Accordingly, the last row of the energy matrix $\varepsilon$ has
the following non-zero entries
\begin{equation} \label{4.1}
\begin{split}
& \varepsilon_{72}=-\frac{45}{2}V &&
\varepsilon_{73}=\frac{441}{4}V && \varepsilon_{74}=-203V \\
& \varepsilon_{75}=\frac{735}{4}V &&
\varepsilon_{76}=-\frac{175}{2}V && \varepsilon_{77}=-\mu+21V
\end{split}
\end{equation}
The matrix $\Omega $ has the expression
\begin{equation}
\Omega =\left(
\begin{array}{ccccccc}
1 & {2^{6}} & {1} & {(2/3)^{6}} & {(1/2)^{6}} & {(2/5)^{6}}
& {(1/3)^{6}} \\
0 & {2}^{5} & {1} & {(2/3)^{5}} & {(1/2)^{5}} & {(2/5)^{5}} & {{{
\ (1/3)^{5}}}} \\
0 & {2^{4}} & {1} & {(2/3)^{4}} & {(1/2)^{4}} & {(2/5)^{4}}
& {(1/3)^{4}} \\
0 & {2^{3}} & {1} & {(2/3)^{3}} & {(1/2)^{3}} & {(2/5)^{3}}
& {(1/3)^{3}} \\
0 & {2^{2}} & {1} & {(2/3)^{2}} & {(1/2)^{2}} & {(2/5)^{2}}
& {(1/3)^{2}} \\
0 & {2} & {1} & {(2/3)} & {(1/2)} & {(2/5)} & {(1/3)} \\
0 & 1 & 1 & 1 & 1 & 1 & 1
\end{array}
\right)  \label{4.3}
\end{equation}
and the normalization matrix $I$ reads as
\begin{equation} \label{4.4}
I=\left(
\begin{array}{ccccccc}
{I_{11}} & {I_{12}} & {I_{13}} & {I_{14}} & {I_{15}} & {I_{16}} & {I_{17}} \\
{I_{12}} & {I_{13}} & {I_{14}} & {I_{15}} & {I_{16}} & {I_{17}} & {I_{27}} \\
{I_{13}} & {I_{14}} & {I_{15}} & {I_{16}} & {I_{17}} & {I_{27}} & {I_{37}} \\
{I_{14}} & {I_{15}} & {I_{16}} & {I_{17}} & {I_{27}} & {I_{37}} & {I_{47}} \\
{I_{15}} & {I_{16}} & {I_{17}} & {I_{27}} & {I_{37}} & {I_{47}} & {I_{57}} \\
{I_{16}} & {I_{17}} & {I_{27}} & {I_{37}} & {I_{47}} & {I_{57}} & {I_{67}} \\
{I_{17}} & {I_{27}} & {I_{37}} & {I_{47}} & {I_{57}} & {I_{67}} &
{I_{77}}
\end{array}
\right)
\end{equation}
where
\begin{align}
&I_{1,p}=\kappa^{(p-1)} \label{4.5} \\
&I_{p,7}=\sum_{m=1}^{6}I_{1,m+1}A_{m}^{(p+5)} \label{4.6} \\
&\kappa^{(p)} = \langle [n^{\alpha}(i)]^{p} \rangle  \label{4.6bis}
\end{align}
Then, the spectral density matrices $\sigma ^{(m)}$ can be easily
computed by means of Eq.~(\ref{3.14}) and the correlation functions
by means of Eq.~(\ref{3.19}).

It is worth noticing that, according to the structure of the
normalization matrix $I$ [(\ref{4.4}) and (\ref{4.6})], which is
dictated by the relation (\ref{3.3}), there exist only seven
independent spectral density matrices $\sigma_{1,p}^{(m)}$ and seven
independent two-point correlation functions $C_{1,p}=\langle c(i)
c^{\dagger }(i) [n^{\alpha }(i)]^{p-1}\rangle$. All others can be
obtained as combinations of these latter according to (\ref{3.3})
and (\ref{4.6}).

As regards the spectral density matrices, we have
\begin{equation}  \label{4.7}
\sigma^{(m)}=\Sigma_{m}\Gamma^{(m)}
\end{equation}
where $\Sigma_{m}$ are functions of the elements $I_{1,p}$ with
$p=1,\ldots,7$ and $\Gamma^{(m)}$ are numerical matrices. In
particular, we have the expressions
\begin{small}
\begin{equation}\label{4.8}
\begin{split}
&\Sigma_{1}=I_{1,1}\\
&+\frac{1}{90}(-441I_{1,2}+812I_{1,3}-735I_{1,4}+350I_{1,5}-84I_{1,6}+8I_{1,7})\\
&\Sigma_{2}=\frac{2}{15}(90I_{1,2}-261I_{1,3}+290I_{1,4}-155I_{1,5}+40I_{1,6}-4I_{1,7})\\
&\Sigma_{3}=\frac{1}{6}(-90I_{1,2}+351I_{1,3}-461I_{1,4}+274I_{1,5}-76I_{1,6}+8I_{1,7})\\
&\Sigma_{4}=\frac{4}{9}(30I_{1,2}-127I_{1,3}+186I_{1,4}-121I_{1,5}+36I_{1,6}-4I_{1,7})\\
&\Sigma_{5}=\frac{1}{6}(-45I_{1,2}+198I_{1,3}-307I_{1,4}+214I_{1,5}-68I_{1,6}+8I_{1,7})\\
&\Sigma_{6}=\frac{2}{15}(18I_{1,2}-81I_{1,3}+130I_{1,4}-95I_{1,5}+32I_{1,6}-4I_{1,7})\\
&\Sigma_{7}=\frac{1}{90}(-30I_{1,2}+137I_{1,3}-225I_{1,4}+170I_{1,5}-60I_{1,6}+8I_{1,7})
\end{split}
\end{equation}
\end{small}
and
\begin{small}
\begin{equation}\label{4.9}
\begin{split}
&\Gamma_{1,m}^{(1)}=\left(1 \; 0 \; 0 \; 0 \; 0 \; 0 \; 0\right)\\
&\Gamma_{1,m}^{(2)}=\left(1 \; 2^{-1}\; 2^{-2} \; 2^{-3} \; 2^{-4}
\; 2^{-5} \; 2^{-6}\right)\\
&\Gamma_{1,m}^{(3)}=\left(1 \; 1 \; 1 \; 1 \; 1 \; 1 \; 1\right)\\
&\Gamma_{1,m}^{(4)}=\left(1 \; (2/3)^{-1} \; (2/3)^{-2} \;
(2/3)^{-3} \; (2/3)^{-4} \; (2/3)^{-5} \; (2/3)^{-6}\right)\\
&\Gamma_{1,m}^{(5)}=\left(1 \; (1/2)^{-1} \; (1/2)^{-2} \;
(1/2)^{-3} \; (1/2)^{-4} \; (1/2)^{-5} \; (1/2)^{-6}\right)\\
&\Gamma_{1,m}^{(6)}=\left(1 \; (2/5)^{-1} \; (2/5)^{-2} \;
(2/5)^{-3} \; (2/5)^{-4} \; (2/5)^{-5} \; (2/5)^{-6}\right)\\
&\Gamma_{1,m}^{(7)}=\left(1 \; (1/3)^{-1} \; (1/3)^{-2} \;
(1/3)^{-3} \; (1/3)^{-4} \; (1/3)^{-5} \; (1/3)^{-6}\right)
\end{split}
\end{equation}
\end{small}

As regards the two-point correlation function, we have
\begin{equation}\label{4.10}
C(i,j)=\delta_{\mathbf{ij}}\frac{1}{2}\sum_{m=1}^{7}T_{m}\sigma
^{(m)}e^{-\mathrm{i}E_{m}(t_{i}-t_{j})}
\end{equation}

\section{Self-consistent equations}

The correlation functions $C_{1,k}$ depend, through the spectral
density matrices that are functions of the entries of the
normalization matrix (\ref{3.14}), on the correlators $\kappa
^{(p)}= \langle [n^{\alpha }(i)]^{p} \rangle$ with $p=1,\ldots,6$.
At this stage, the correlators $\kappa ^{(p)}$ are unknown as the
operators $[n^{\alpha }(i)]^{p}$ do not belong to the chosen basis
(\ref{3.5}).

To the purpose of definitively computing the correlation functions
$C_{1,k}$, let us introduce the following operators
\begin{equation} \label{5.3}
\begin{split}
&\xi _{a}(i)=[1-n(i)+D(i)]c_{a}(i) \\
&\eta _{a}(i)=[n(i)-2D(i)]c_{a}(i) \\
&\zeta _{a}(i)=D(i)c_{a}(i)
\end{split}
\end{equation}
where the double $D(i)$ and the triple $T(i)$ occupancy operators at
the site $\mathbf{i}$ read as
\begin{equation}\label{5.2}
\begin{split}
&D(i)=\sum_{\substack{a,b=1\\a<b}}^3
n_{a}(i)n_{b}(i) \\
&T(i)=\prod_{a=1}^3n_{a}(i)
\end{split}
\end{equation}
The operators $\xi_{a}(i)$, $\eta_{a}(i)$ and $\zeta_{a}(i)$ rule
the transitions among states with different particle number at the
site $\mathbf{i}$: $0 \Leftrightarrow 1$, $1 \Leftrightarrow 2$, $2
\Leftrightarrow 3$, respectively. Their complex algebra is presented
in detail in Appendix B. In particular, they satisfy the following
relevant relations
\begin{equation} \label{5.4}
\begin{split}
& c_{a}(i)=\xi _{a}(i)+\eta _{a}(i)+\zeta_{a}(i) \\
&
\xi^{\dagger}_{a}(i)\eta_{b}(i)=\xi^{\dagger}_{a}(i)\zeta_{b}(i)=\eta^{\dagger
}_{a}(i)\zeta_{b}(i)=0
\end{split}
\end{equation}
and
\begin{equation} \label{5.4bis}
\begin{split}
& \xi_{a}^{\dagger}(i)n(i)=0 && \xi _{a}^{\dagger}(i)D(i)=0 &&
\xi_{a}^{\dagger}(i)T(i)=0 \\
& \eta_{a}^{\dagger}(i)n(i)=\eta_{a}^{\dagger}(i) &&
\eta_{a}^{\dagger}(i)D(i)=0 && \eta_{a}^{\dagger}(i)T(i)=0 \\
& \zeta_{a}^{\dagger}(i)n(i)=2\zeta_{a}^{\dagger}(i) &&
\zeta_{a}^{\dagger}(i)D(i)=\zeta_{a}^{\dagger}(i) &&
\zeta_{a}^{\dagger}(i)T(i)=0
\end{split}
\end{equation}
for any choice of the indices $a$ and $b$.

After (\ref{5.4bis}), it is possible to show that (see Appendix C)
\begin{equation} \label{5.5}
\begin{split}
\xi^{\dagger }(i)e^{-\beta H} & =\xi^{\dagger }(i)e^{-\beta H_{0i}} \\
\eta^{\dagger }(i)e^{-\beta H} & =\eta^{\dagger}(i)
\left\{1+\sum_{m=1}^{6}f_{m}[n^{\alpha}(i)]^{m}\right\}e^{-\beta H_{0i}} \\
\zeta^{\dagger }(i)e^{-\beta H} & =\zeta ^{\dagger
}(i)e^{-\beta H_{0i}}\\
&+ \zeta ^{\dagger }(i) \sum_{m=1}^{6}(2f_{m}+g_{m})[n^{\alpha
}(i)]^{m} e^{-\beta H_{0i}}
\end{split}
\end{equation}
where $H_{0i}$ is a reduced part of the Hamiltonian
\begin{equation} \label{5.6}
\begin{split}
& H_{0i} = H - H_{Ii} \\
& H_{Ii} = 2V n(i) n^{\alpha }(i)
\end{split}
\end{equation}
We have used for $\xi(i)$, $\eta(i)$ and $\zeta(i)$ the same
vectorial notation that has been used for $c(i)$ (see (\ref{2.3})).
The coefficients $f_{m}$ and $g_{m}$ are defined in Appendix C. We
will exploit the relations (\ref{5.5}) in order to compute the
correlation functions
\begin{equation} \label{5.7}
\begin{split}
& C_{1,k}^{\xi\xi} = \langle \xi(i) \xi^{\dagger }(i) [n^{\alpha }(i)]^{k-1} \rangle \\
& C_{1,k}^{\eta\eta} = \langle \eta(i) \eta^{\dagger }(i) [n^{\alpha }(i)]^{k-1} \rangle \quad \quad k=1,\ldots,7 \\
& C_{1,k}^{\zeta\zeta} = \langle \zeta(i)\zeta ^{\dagger}(i)
[n^{\alpha }(i)]^{k-1} \rangle
\end{split}
\end{equation}
that, after the relations (\ref{5.4}), sum up to
$C_{1,k}=C_{1,k}^{\xi\xi}+C_{1,k}^{\eta\eta}+C_{1,k}^{\zeta\zeta}$.
By means of (\ref{5.5}), we have
\begin{equation} \label{5.8}
\begin{split}
C_{1,k}^{\xi\xi}& =\frac{\langle \xi(i) \xi^{\dagger}(i)
[n^{\alpha}(i)]^{k-1}\rangle_{0i}}{\langle e^{-\beta H_{Ii}} \rangle_{0i}} \\
C_{1,k}^{\eta\eta}& =\frac{\langle \eta (i)\eta ^{\dagger
}(i)[n^{\alpha }(i)]^{k-1}\rangle _{0i}}{\langle e^{-\beta H_{Ii}}\rangle _{0i}} \\
& +\sum_{m=1}^{6}f_{m} \frac{\langle \eta (i)\eta ^{\dagger
}(i)[n^{\alpha }(i)]^{k-1+m}\rangle _{0i}}{\langle e^{-\beta H_{Ii}}\rangle _{0i}} \\
C_{1,k}^{\zeta\zeta}& = \frac{\langle \zeta (i)\zeta ^{\dagger
}(i)[n^{\alpha }(i)]^{k-1}\rangle _{0i}}{\langle e^{-\beta H_{Ii}}\rangle _{0i}}\\
& + \sum_{m=1}^{6}(2f_{m}+g_{m}) \frac{\langle \zeta (i)\zeta
^{\dagger }(i)[n^{\alpha }(i)]^{k-1+m}\rangle _{0i}}{\langle
e^{-\beta H_{Ii}}\rangle _{0i}}
\end{split}
\end{equation}
where for a generic operator $O$ the notation $\langle O\rangle
_{0i}$ denotes the thermal average with respect to $H_{0i}$
\begin{equation} \label{5.9}
\langle O\rangle _{0i}={\frac{{Tr\{Oe^{-\beta
H_{0i}}\}}}{{Tr\{e^{-\beta H_{0i}}\}}}}
\end{equation}

Now, we observe that $H_{0i}$ describes a system where the site
$\mathbf{i}$ is not connected to any other site of the chain. Then,
at the site $\mathbf{i}$ the local operators enjoy a free dynamics
\begin{equation} \label{5.10}
\begin{split}
[\xi(i),H_{0i}] = -\mu\xi(i) \\
[\eta(i),H_{0i}] = -\mu\eta(i) \\
[\zeta(i),H_{0i}] = -\mu\zeta(i)
\end{split}
\end{equation}
By means of these equations of motion and by making use of the
relations (\ref{B.8}), it is possible to derive
\begin{equation}\label{5.11}
\begin{split}
& \left\langle \xi(i) \xi^\dag(i) [n^\alpha
(i)]^{k-1}\right\rangle_{0i}=\frac{1}{(1+e^{\beta
\mu })^3}\left\langle[n^\alpha(i)]^{k-1}\right\rangle_{0i} \\
& \left\langle \eta(i) \eta^\dag(i)
[n^\alpha(i)]^{k-1}\right\rangle_{0i}=\frac{2e^{\beta \mu }
}{(1+e^{\beta \mu })^3}\left\langle[n^\alpha(i)]^{k-1}\right\rangle_{0i} \\
& \left\langle \zeta(i) \zeta^\dag(i) [n^{\alpha
}(i)]^{k-1}\right\rangle_{0i}=\frac{e^{2\beta \mu }}{(1+e^{\beta \mu
})^3}\left\langle[n^\alpha (i)]^{k-1}\right\rangle_{0i}
\end{split}
\end{equation}
and rewrite (\ref{5.8}) as
\begin{equation}\label{5.12}
\begin{split}
C_{1,k}^{\xi \xi }& =\frac{1}{(1+e^{\beta \mu })^{3}}\frac{\langle
[n^{\alpha}(i)]^{k-1}\rangle_{0i}}{\langle e^{-\beta H_{Ii}}\rangle _{0i}} \\
C_{1,k}^{\eta \eta }& =\frac{2e^{\beta \mu }}{(1+e^{\beta \mu
})^{3}} \frac{\langle [n^{\alpha }(i)]^{k-1}\rangle
_{0i}}{\langle e^{-\beta H_{Ii}}\rangle _{0i}} \\
& +\frac{2e^{\beta \mu }}{(1+e^{\beta \mu })^{3}}
\sum_{m=1}^{6}f_{m} \frac{\langle
[n^{\alpha}(i)]^{m+k-1}\rangle _{0i}}{\langle e^{-\beta H_{Ii}}\rangle _{0i}} \\
C_{1,k}^{\zeta \zeta }& =\frac{e^{2\beta \mu }}{(1+e^{\beta \mu
})^{3}}\frac{\langle [n^{\alpha }(i)]^{k-1}\rangle
_{0i}}{\langle e^{-\beta H_{Ii}}\rangle _{0i}}\\
& +\frac{e^{2\beta \mu }}{(1+e^{\beta \mu
})^{3}}\sum_{m=1}^{6}(2f_{m}+g_{m})\frac{\langle [n^{\alpha
}(i)]^{m+k-1}\rangle _{0i}}{\langle e^{-\beta H_{Ii}}\rangle _{0i}}
\end{split}
\end{equation}
Therefore, we are left with the problem of computing the functions
$\left\langle[n^\alpha(i)]^p\right\rangle_{0i}$.

Now, we observe that $H_{0i}$ describes a system where the original
infinite chain is split into two disconnected infinite sub-chains
(the infinite chains to the left and to the right of the site
$\mathbf{i}$) plus the site $\mathbf{i}$. Then, in the
$H_{0i}$-representation, correlation functions which relate sites
belonging to different sub-chains and/or the site $\mathbf{i}$, can
be decoupled:
\begin{equation} \label{5.13}
\langle a(j) b(l)\rangle _{0i}= \langle a(j) \rangle _{0i} \langle
b(l) \rangle _{0i}
\end{equation}
when, for instance, $\mathbf{j}\leq\mathbf{i}\leq\mathbf{l}$. $a(j)$
and $b(l)$ are any functions of $n(j)$ and $n(l)$, respectively. Let
us recall Eqs.~(\ref{A.8}) and (\ref{A.9}). By using the property
(\ref{5.13}), we have
\begin{equation}\label{5.14}
\begin{split}
& \langle Z_{0}(i) \rangle_{0i} = 2X_{1} \\
& \langle Z_{1}(i) \rangle_{0i} = 2X_{2}+X_{1}^{2} \\
& \langle Z_{2}(i) \rangle_{0i} = 2X_{3}+2X_{1}X_{2} \\
& \langle Z_{3}(i) \rangle_{0i} = 2X_{1}X_{3}+X_{2}^{2} \\
& \langle Z_{4}(i) \rangle_{0i} = 2X_{2}X_{3} \\
& \langle Z_{5}(i) \rangle_{0i} = X_{3}^{2}
\end{split}
\end{equation}
with
\begin{equation}\label{5.15}
\begin{split}
& X_{1} = \langle n^{\alpha}(i) \rangle_{0i} \\
& X_{2} = \langle D^{\alpha}(i) \rangle_{0i} \\
& X_{3} = \langle T^{\alpha}(i) \rangle_{0i}
\end{split}
\end{equation}
Therefore, we have
\begin{equation}\label{5.16}
\begin{split}
&\langle [n^{\alpha }(i)]^{p}\rangle _{0i} = \frac{1}{2^{p}}
\left[2X_{1}+b_{1}^{(p)}(2X_{2}+X_{1}^{2}) \right.\\
&\left. +2b_{2}^{(p)}(X_{3}+X_{1}X_{2})
+b_{3}^{(p)}(2X_{1}X_{3}+X_{2}^{2}) \right.\\
&\left. +2b_{4}^{(p)}X_{2}X_{3}+b_{5}^{(p)}X_{3}^{2}\right]
\end{split}
\end{equation}
The problem of computing all two-point correlation functions is thus
reduced to the problem of computing just three parameters: $X_{1}$,
$X_{2}$ and $X_{3}$. If we suppose the system to be homogenous, we
can use the following three self-consistent equations in order to
determine the unknown parameters
\begin{equation} \label{5.17}
\begin{split}
& \langle n(i) \rangle = \langle n^{\alpha }(i) \rangle \Rightarrow \langle n(i)e^{-\beta H_{1}}\rangle _{0i}=\langle n^{\alpha }(i)e^{-\beta H_{1}}\rangle _{0i}\\
& \langle D(i) \rangle = \langle D^{\alpha }(i) \rangle \Rightarrow \langle D(i)e^{-\beta H_{1}}\rangle _{0i}=\langle D^{\alpha }(i)e^{-\beta H_{1}}\rangle _{0i}\\
& \langle T(i) \rangle = \langle T^{\alpha }(i) \rangle \Rightarrow
\langle T(i)e^{-\beta H_{1}}\rangle _{0i}=\langle T^{\alpha
}(i)e^{-\beta H_{1}}\rangle _{0i}
\end{split}
\end{equation}
By means of Eqs.~(\ref{C.3}) and (\ref{5.19}), we have
\begin{equation} \label{5.20}
\begin{split}
&\left \langle n(i) e^{-\beta H_{Ii}} \right\rangle _{0i} = B_1 +
\sum_{m=1}^{6} \left [ f_{m}(B_1+ 2B_2) \right.\\
&\left. + g_{m} (2B_2 + 3B_3)+ h_{m}
3B_3 \right ] \left\langle [n^{\alpha }(i)]^{m} \right\rangle_{0i} \\
&\left\langle n^{\alpha }(i) e^{-\beta H_{Ii}} \right\rangle _{0i} =
X_{1} + \sum _{m=1}^{6} \left[ f_{m} B_1 + g_{m} B_2 \right.\\
&\left. + h_{m} B_3 \right] \left\langle [n^{\alpha }(i)]^{m+1}
\right\rangle _{0i}
\end{split}
\end{equation}
\begin{equation}\label{5.21}
\begin{split}
& \langle D(i) e^{-\beta H_{Ii}} \rangle _{0i} = B_{2} +
\sum_{m=1}^{6} \left[
(2B_{2}+3B_{3})f_{m} \right.\\
&\left. +(B_{2}+6B_{3})g_{m}+3B_{3}h_{m}\right]\langle [n^{\alpha }(i)]^{m} \rangle_{0i} \\
& \langle D^{\alpha}(i) e^{-\beta H_{Ii}} \rangle _{0i} = X_{2} +
\sum_{m=1}^{6} \left[B_{1}f_{m}+B_{2}g_{m}\right.\\
&\left.+B_{3}h_{m}\right]\langle D^{\alpha}(i)[n^{\alpha
}(i)]^{m}\rangle_{0i}
\end{split}
\end{equation}
\begin{equation}\label{5.22}
\begin{split}
& \langle T(i) e^{-\beta H_{Ii}} \rangle _{0i} = B_{3}
\left[1+\sum_{m=1}^{6}\left(3f_{m}+3g_{m}\right.\right.\\
&\left.\left.+h_{m}\right)\right]\langle [n^{\alpha }(i)]^{m} \rangle_{0i} \\
& \langle T^{\alpha }(i) e^{-\beta H_{Ii}} \rangle _{0i} =
X_{3}+\sum_{m=1}^{6}\left(B_{1}f_{m}+B_{2}g_{m}\right.\\
&\left.+B_{3}h_{m}\right)\langle T^{\alpha }(i)[n^{\alpha
}(i)]^{m}\rangle_{0i}
\end{split}
\end{equation}
where
\begin{equation} \label{5.10bis}
\begin{split}
& B_1=\langle n(i) \rangle _{0i} =
\frac{3e^{\beta\mu}}{e^{\beta\mu}+1} \\
& B_2=\langle D(i) \rangle _{0i} = \frac{3e^{2\beta\mu}}{(e^{\beta\mu}+1)^2} \\
& B_3=\langle T(i) \rangle _{0i} =
\frac{e^{3\beta\mu}}{(e^{\beta\mu}+1)^3}
\end{split}
\end{equation}

We need to calculate the averages $\langle D^{\alpha }(i)[n^{\alpha
}(i)]^{m}\rangle_{0i}$ and $\langle T^{\alpha }(i)[n^{\alpha
}(i)]^{m}\rangle_{0i}$. By using (\ref{A.8})
\begin{equation}\label{5.24}
\begin{split}
&\langle D^{\alpha }(i)[n^{\alpha }(i)]^{p}\rangle_{0i} =
\frac{1}{2^{p}}
\sum_{m=0}^{5}b_{m}^{(p)}\langle D^{\alpha }(i) Z_{m}(i)\rangle_{0i} \\
&\langle T^{\alpha }(i)[n^{\alpha }(i)]^{p}\rangle_{0i} =
\frac{1}{2^{p}} \sum_{m=0}^{5} b_{m}^{(p)} \langle T^{\alpha }(i)
Z_{m}(i)\rangle_{0i}
\end{split}
\end{equation}
By recalling the definitions (\ref{A.9}) and using the property
(\ref{5.13})
\begin{equation}\label{5.25}
\begin{split}
&\langle D^{\alpha }(i)[n^{\alpha }(i)]^{p}\rangle_{0i} =
\frac{1}{2^{p}}
\left[2X_{2}+3X_{3}+X_{1}X_{2}\right. \\
& \left.+b_{1}^{(p)}(X_{2}+6X_{3}+X_{2}^{2}+2X_{1}X_{2}+3X_{1}X_{3})
\right. \\
&\left.+b_{2}^{(p)}(3X_{3}+2X_{2}^{2}+X_{1}X_{2}+6X_{1}X_{3}+4X_{2}X_{3})\right. \\
& \left.+b_{3}^{(p)}(3X_{1}X_{3}+X_{2}^{2}+8X_{2}X_{3}+3X_{3}^{2})
\right. \\
&\left.+2b_{4}^{(p)}(2X_{2}X_{3}+3X_{3}^{2})+3b_{5}^{(p)}X_{3}^{2}\right]
\end{split}
\end{equation}
\begin{equation}\label{5.26}
\begin{split}
&\langle T^{\alpha }(i)[n^{\alpha }(i)]^{p}\rangle_{0i}=
\frac{1}{2^{p}} \left[
3X_{3}+X_{1}X_{3}\right. \\
& \left.+b_{1}^{(p)}(3X_{3}+3X_{1}X_{3}+X_{2}X_{3})
\right. \\
& \left.+b_{2}^{(p)}(X_{3}+X_{3}^{2} +3X_{2}X_{3}+3X_{1}X_{3})
\right. \\
& \left.+b_{3}^{(p)}(3X_{2}X_{3}+X_{1}X_{3}+3X_{3}^{2})
\right. \\
&
\left.+b_{4}^{(p)}(3X_{3}^{2}+X_{2}X_{3})+b_{5}^{(p)}X_{3}^{2}\right]
\end{split}
\end{equation}

Summarizing, the three parameters $X_{1}$, $X_{2}$, $X_{3}$ are
determined by the coupled self-consistent equations
(\ref{5.20})-(\ref{5.22}), where the averages $\left\langle
[n^\alpha (i)]^m\right\rangle_{0i}$, $\left\langle D^\alpha
(i)[n^\alpha(i)]^m\right\rangle_{0i}$ and $\left\langle T^\alpha
(i)[n^\alpha(i)]^m\right\rangle_{0i}$ are computed by means of
(\ref{5.16}), (\ref{5.25}) and (\ref{5.26}), respectively. Once we
know the three parameters, we can calculate the correlation
functions and all the properties. For example,
\begin{equation}\label{5.27}
\begin{split}
& C_{1,k}^{\xi\xi}=\frac{1-B_{1}+B_{2}-B_{3}}{\left\langle e^{-\beta
H_I }\right\rangle_{0i}} \left\langle [n^\alpha(i)]^{k-1}\right\rangle _{0i} \\
& C_{1,k}^{\eta \eta }=\frac{2B_{1}-4B_{2}+6B_{3}}{3\left\langle
e^{-\beta H_I }\right\rangle _{0i}}[\left\langle [n^\alpha
(i)]^{k-1}\right\rangle_{0i}\\&+\sum_{m=1}^{6}f_{m}\left\langle
[n^\alpha(i)]^{m+k-1}\right\rangle_{0i}] \\
& C_{1,k}^{\zeta\zeta}=\frac{B_{2}-3B_{3}}{3\left\langle e^{-\beta
H_I }\right\rangle_{0i}}[\left\langle
[n^\alpha(i)]^{k-1}\right\rangle_{0i}\\&+\sum_{m=1}^{6}(2f_{m}+g_{m})\left\langle[n^\alpha
(i)]^{m+k-1}\right\rangle_{0i}]\\
& C_{1,k}=C_{1,k}^{\xi \xi }+C_{1,1}^{\eta \eta }+C_{1,1}^{\zeta
\zeta } \\
& \left\langle n\right\rangle =3-3C_{1,k} \\
& \left\langle D\right\rangle =3-3(C_{1,1}^{\xi \xi
}+\frac{3}{2}C_{1,1}^{\zeta \zeta}+2C_{1,1}^{\eta \eta }) \\
& \left\langle T\right\rangle =1-3(\frac{1}{3}C_{1,1}^{\xi \xi
}+\frac{1}{2}C_{1,1}^{\zeta \zeta }+C_{1,1}^{\eta \eta })
\\
&\kappa ^{(p)}=\frac{\left\langle [n^\alpha(i)]^p
\right\rangle_{0i}}{\left\langle e^{-\beta H_I}\right\rangle_{0i}
}\\
&+\sum_{m=1}^{6}(B_{1}f_{m}+B_{2}g_{m}+B_{3}h_{m})\frac{\left\langle
[n^\alpha(i)]^{m+p}\right\rangle_{0i}}{\left\langle e^{-\beta H_I
}\right\rangle _{0i}}
\end{split}
\end{equation}
\begin{equation}
\begin{split}
& \left\langle n(i)[n^\alpha(i)]^p \right\rangle =
B_{1}\frac{\left\langle [n^\alpha
(i)]^p\right\rangle_{0i}}{\left\langle e^{-\beta H_I }\right\rangle
_{i0}}\\&+\sum_{m=1}^{6}\left[(B_{1}+2B_{2})f_{m}+(2B_{2}+3B_{3})g_{m}\right. \\
& \left.+3B_{3}h_{m}\right]\frac{\left\langle [n^\alpha
(i)]^{m+p}\right\rangle_{0i}}{\left\langle e^{-\beta H_I
}\right\rangle_{0i}}
\end{split}
\end{equation}
The coefficients $f_{m}$, $g_{m}$ and $h_{m}$ are defined in
Appendix C. The average $\left\langle e^{-\beta H_I
}\right\rangle_{0i}$ can be computed by means of (\ref{C.3}) and has
the expression
\begin{equation}\label{5.31}
\left\langle e^{-\beta H_I
}\right\rangle_{0i}=1+\sum_{m=1}^{6}(B_{1}f_{m}+B_{2}g_{m}+B_{3}h_{m})\left\langle
[n^\alpha(i)]^m\right\rangle_{0i}
\end{equation}

\section{Results}

\begin{figure}[tbp]
\includegraphics[width=7cm]{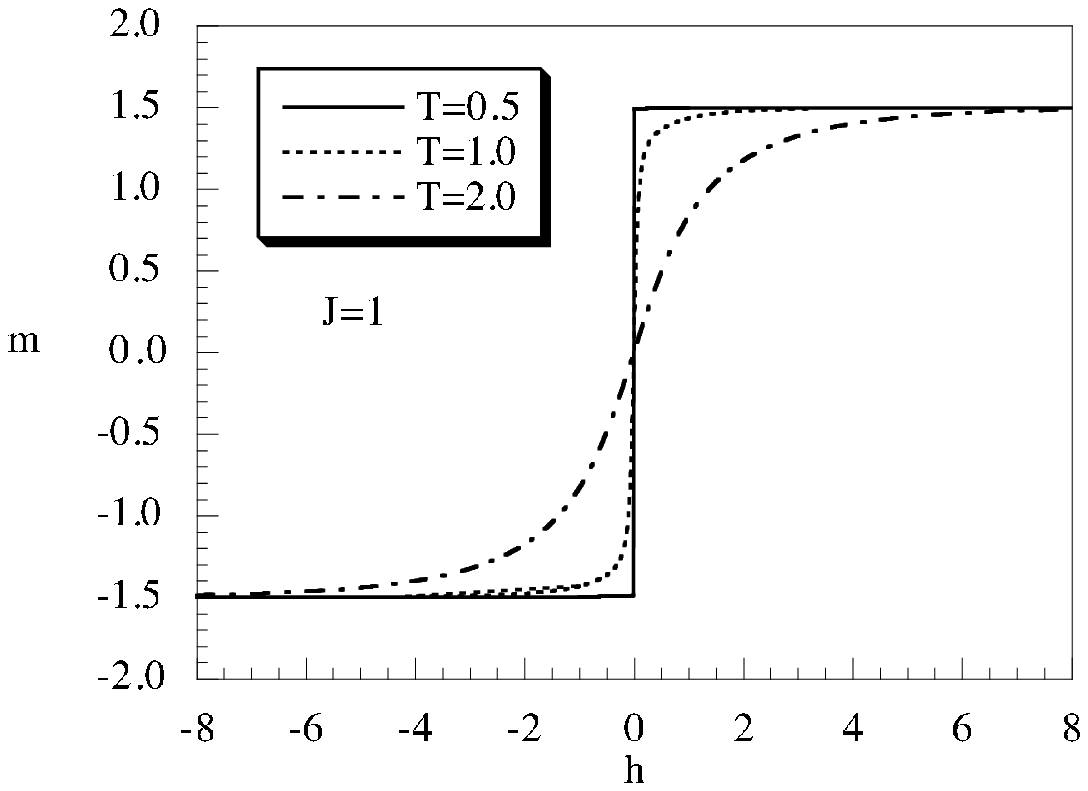}
\includegraphics[width=7cm]{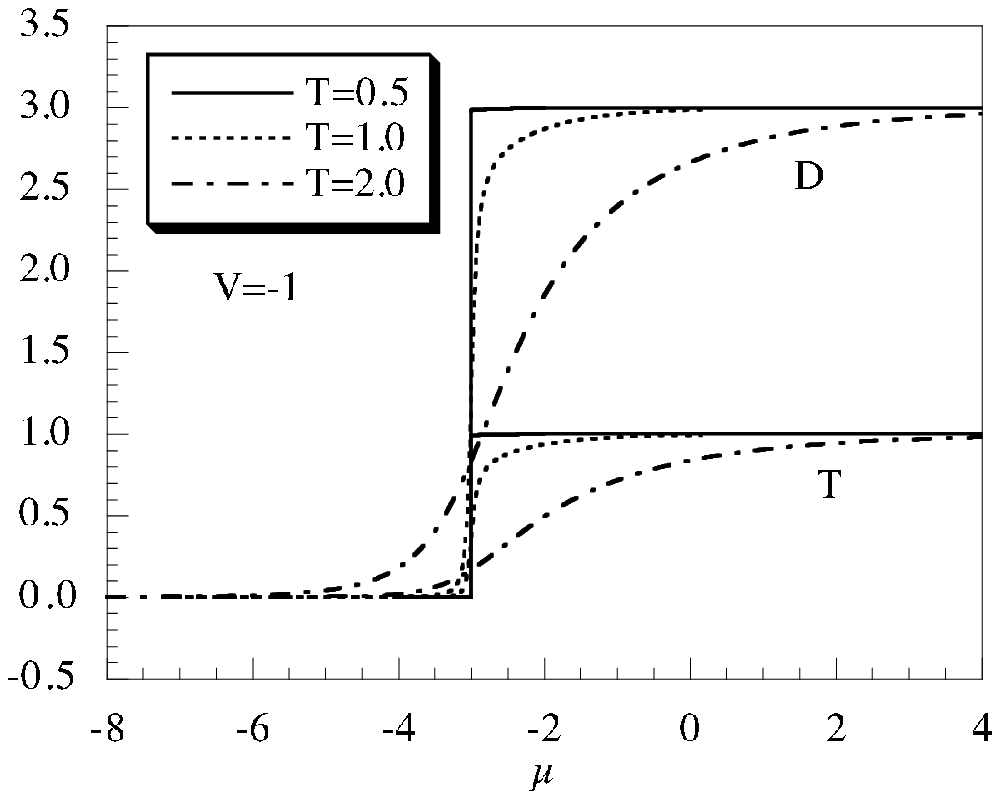}
\includegraphics[width=7cm]{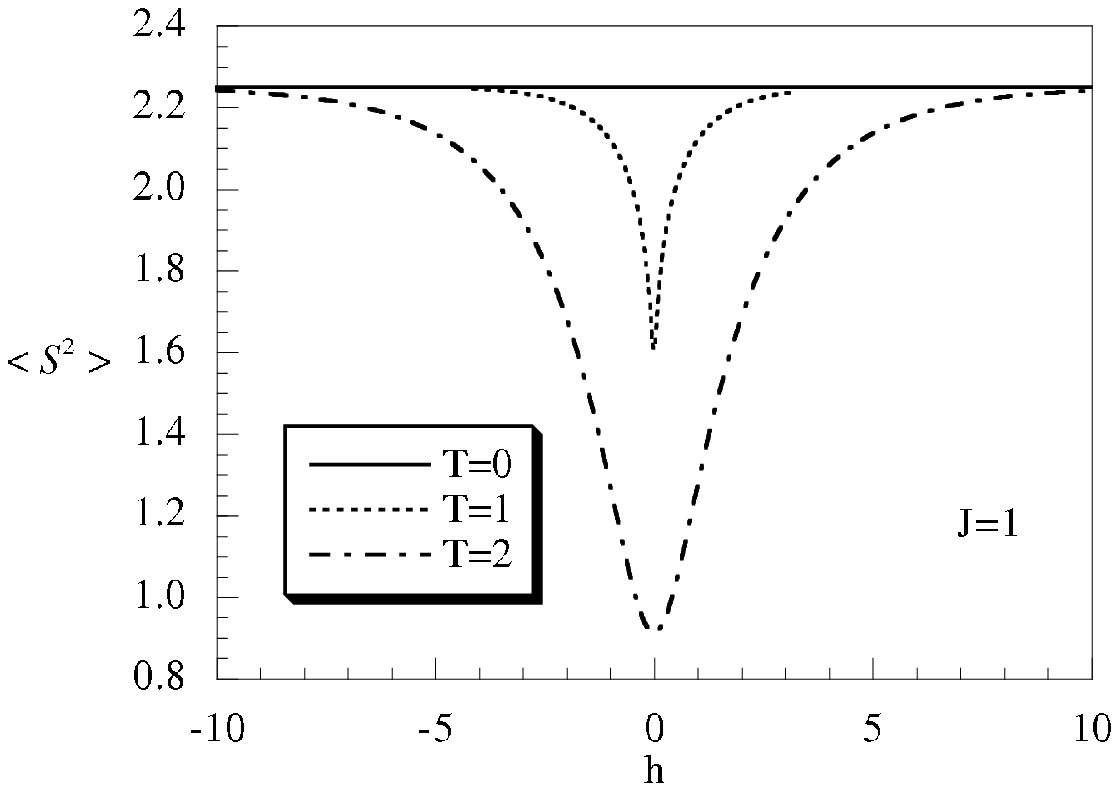}
\includegraphics[width=7cm]{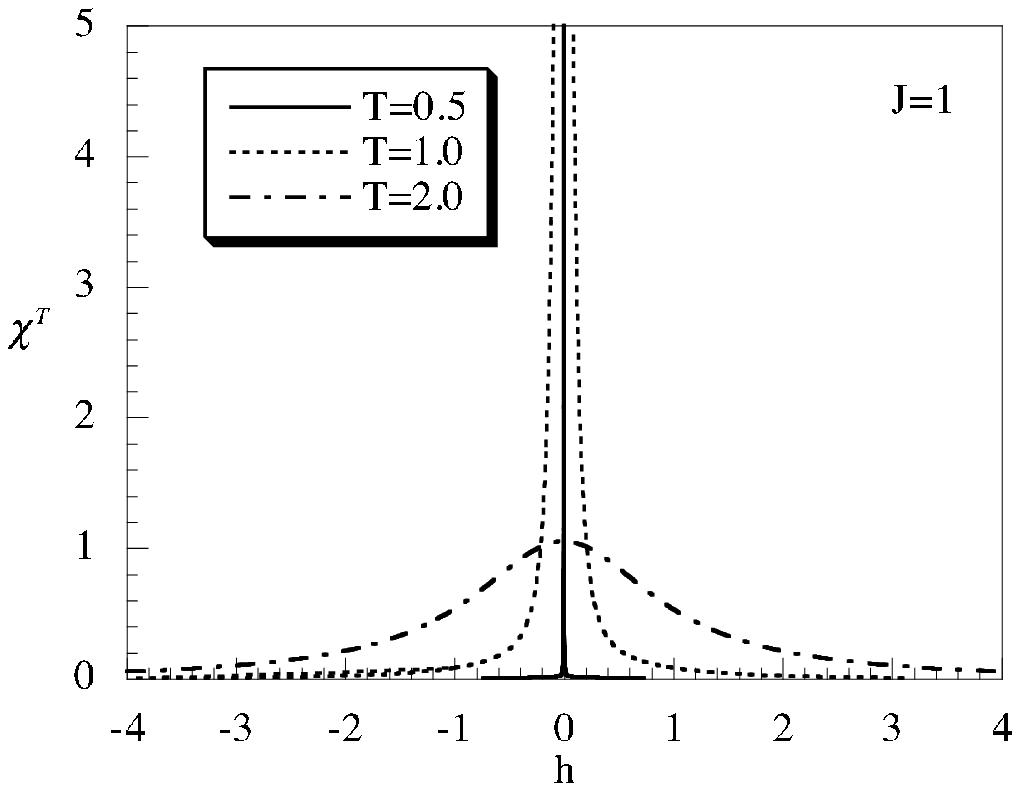}
\caption{(top) Magnetization per site $m$ as a function of the
magnetic field $h$ at $J=1$. (middle top) Double $D$ and triple $T$
occupancies per site as functions of the chemical potential $\mu$ at
$V=-1$. (middle bottom) Square magnetic moment $\langle S^2 \rangle$
per site as function of the magnetic field $h$ at $J=1$. (bottom)
Magnetic susceptibility $\chi$ as a function of the magnetic field
$h$ at $J=1$. In all panels, $T=0$ (only in middle bottom), $0.5$
(except for middle bottom), $1$ and $2$.} \label{Fig.1}
\end{figure}

\begin{figure}[tbp]
\includegraphics[width=7cm]{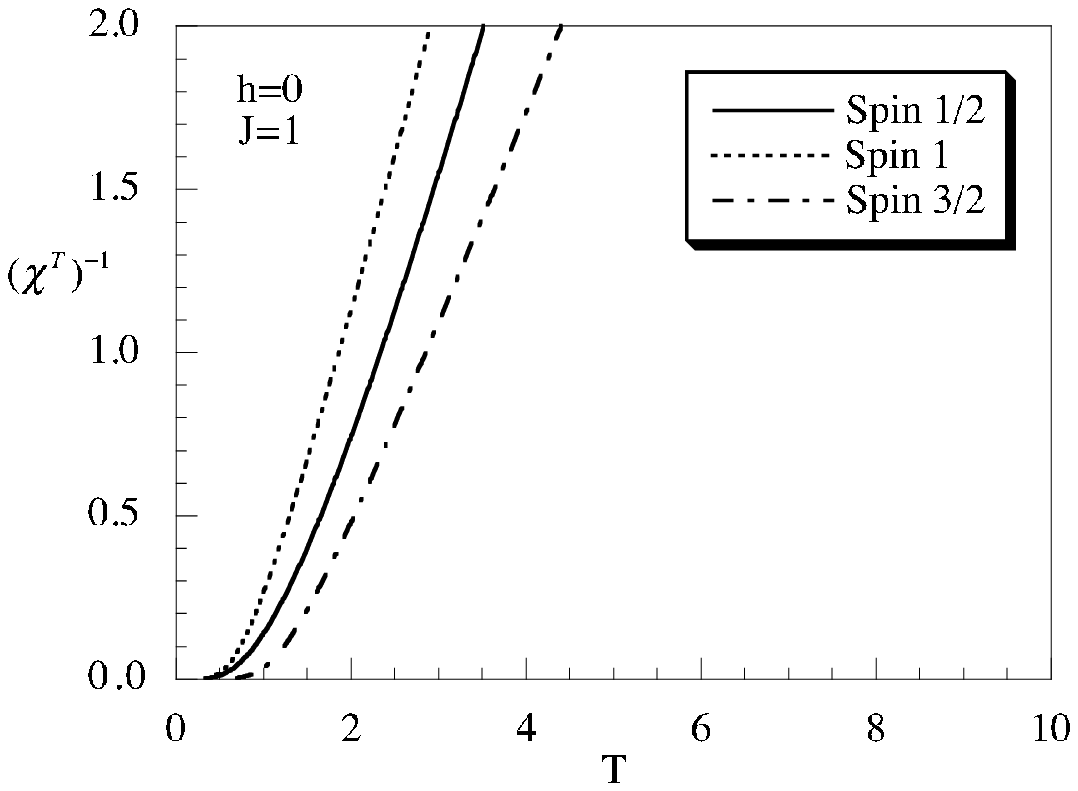}
\includegraphics[width=7cm]{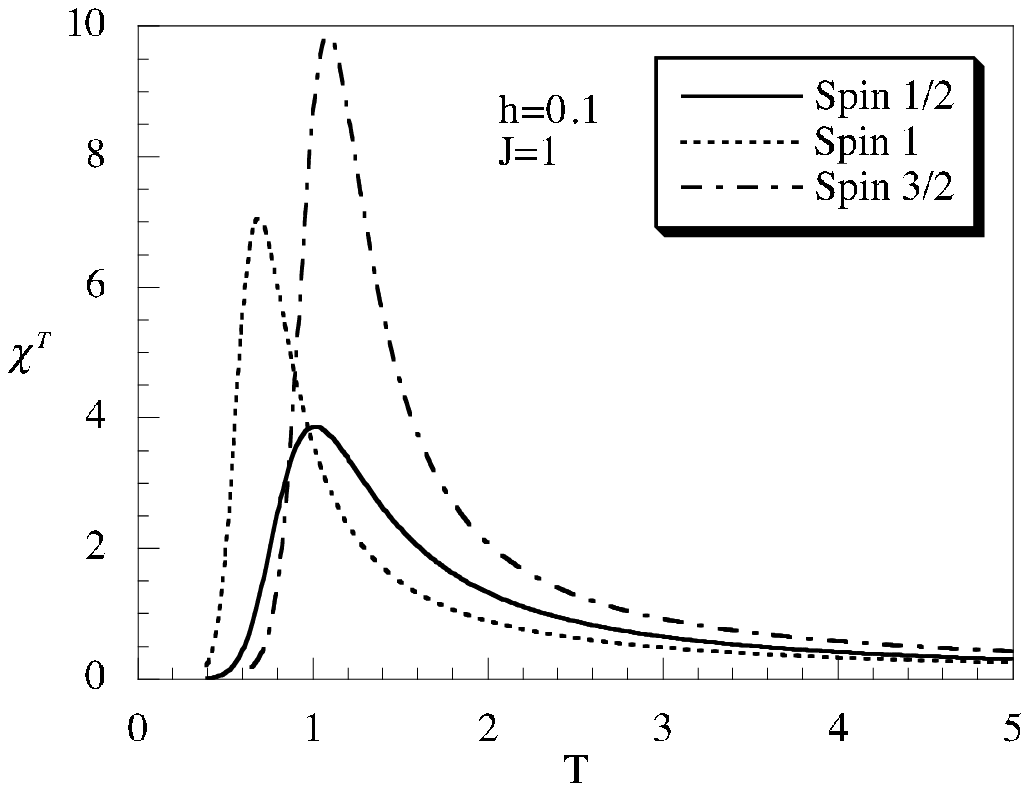}
\caption{(top) Inverse magnetic susceptibility $\chi^{-1}$ at $h=0$
and (bottom) magnetic susceptibility $\chi$ at $h=0.1$ as functions
of the temperature $T$ at $J=1$ and spin $1/2$, $1$ and $3/2$.}
\label{Fig.2}
\end{figure}

\begin{figure}[tbp]
\includegraphics[width=7cm]{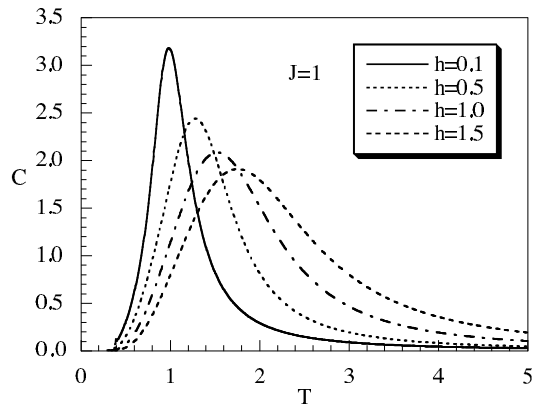}
\caption{Specific heat $C$ as a function of the temperature $T$ at
$J=1$ and $h=0.1$, $0.5$, $1$ and $1.5$.} \label{Fig.3}
\end{figure}

\subsection{Ferromagnetic coupling  $J=1$}

In Fig.~\ref{Fig.1}, we report the magnetization per site $m$ as a
function of the magnetic field $h$ at $J=1$, the double $D$ and the
triple $T$ occupancies per site as functions of the chemical
potential $\mu$ at $V=-1$, the square magnetic moment $\langle S^2
\rangle$ per site as function of the magnetic field $h$ at $J=1$ and
the magnetic susceptibility $\chi$ as a function of the magnetic
field $h$ at $J=1$ and $T=0$ (only in middle bottom panel), $0.5$
(except for middle bottom panel), $1$ and $2$. At the lower
temperature ($T=0.5$), going from negative to positive values of the
magnetic field, the magnetization jumps from $-\frac32$ to $\frac32$
and the susceptibility diverges exactly at zero field. This is the
signature of a quite sharp transition between spin configurations
with maximum possible magnetization and direction dictated by the
value of the external magnetic field. This immediately reflects on
the values of the double and triple occupancies that quite sharply
jumps from zero to their maximum possible values: $3$ and $1$,
respectively. The temperature just makes the transition smoother and
forces the system to pass through excited states with a lower value
of the absolute magnetization as it results clear by looking at the
values of the square magnetic moment.

In Fig.~\ref{Fig.2}, we report the inverse magnetic susceptibility
$\chi^{-1}$ at $h=0$ and the magnetic susceptibility $\chi$ at
$h=0.1$ as functions of the temperature $T$ at $J=1$ and spin $1/2$,
$1$ and $3/2$. The sharp transition at zero temperature in zero
external magnetic field is here clearly signalled by the divergence
of the magnetic susceptibility for any value of the spin. At finite
external magnetic field $h=0.1$, the susceptibility presents a
maximum at a finite temperature whose value depends on the value of
the spin. Both at very low and very high temperatures, the
magnetization stays constant at its maximum value or at zero,
respectively, almost independently on the value of the external
magnetic field. According to this, in these two regimes the
susceptibility is zero or practically zero. At intermediate
temperatures, the value of the magnetization definitely depends on
the value of the external magnetic field. This explains the
bell-shape of the susceptibility curve and the presence of a
maximum.

In Fig.~\ref{Fig.3}, we report the specific heat $C$ as a function
of the temperature $T$ at $J=1$ and $h=0.1$, $0.5$, $1$ and $1.5$.
The exponential behavior at low temperatures clearly signals the
presence of at least one gap in the excitation spectrum between the
fully polarized state, which is the ground state for any non-zero
value of the external magnetic field, and the first excited state to
which corresponds a lower value of the magnetization. This behavior
also explains why we have been able to see a sharp transition in the
magnetization as a function of the external magnetic field at
temperatures as high as $T=0.5$. The temperature $T_M$, at which is
clearly observable a maximum in all curves, strongly depends on the
value of the external magnetic field $h$. In particular, $T_M$
increases on increasing $h$: the gap between the ground state and
the first excited state also increase on increasing $h$. We can
speculate that $T_M \approx \frac12(h+2J)$, that is half of the gap
existing between the fully polarized states with spin $\frac32$ and
$\frac12$.

\begin{figure}[tbp]
\includegraphics[width=7cm]{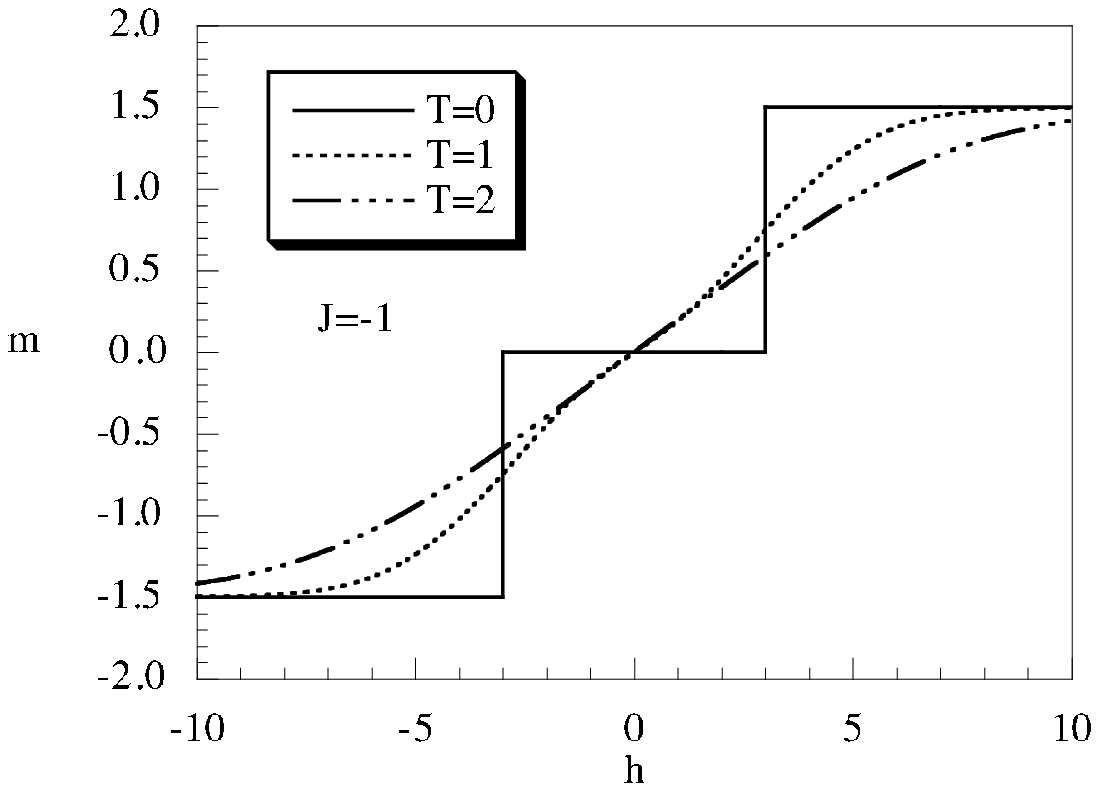}
\includegraphics[width=7cm]{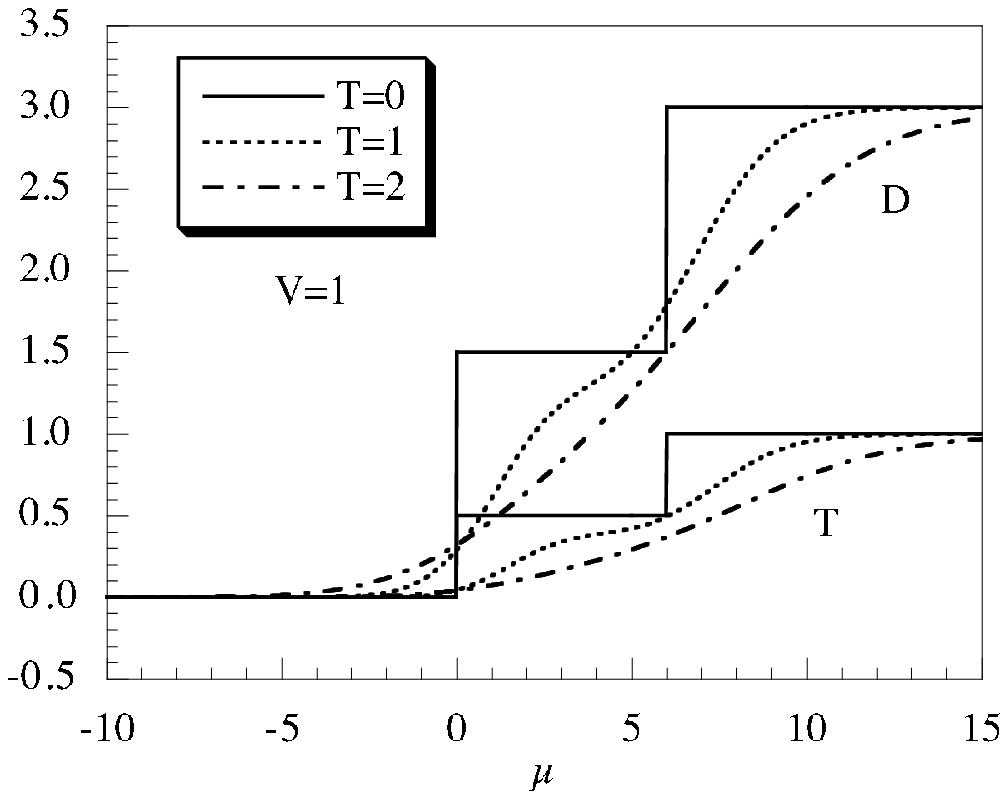}
\includegraphics[width=7cm]{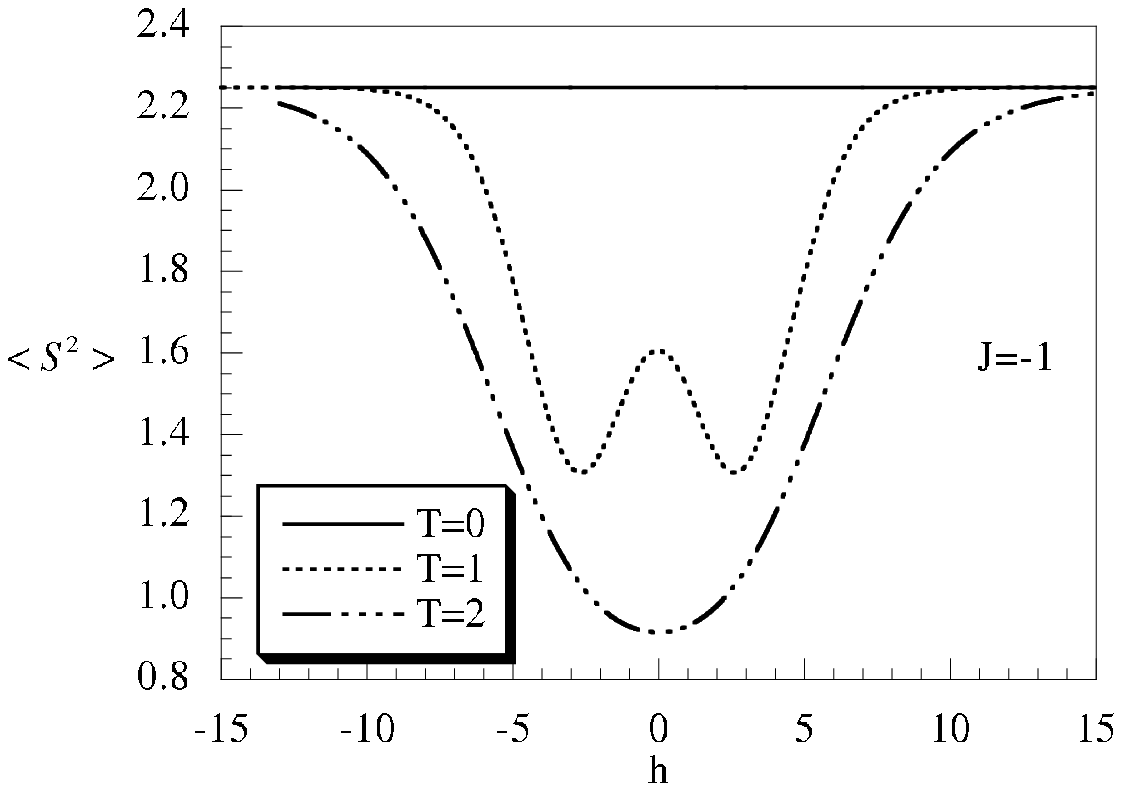}
\caption{(top) Magnetization per site $m$ as a function of the
magnetic field $h$ at $J=-1$. (middle) Double $D$ and triple $T$
occupancies per site as functions of the chemical potential $\mu$ at
$V=1$. In both panels, $T=0$, $1$ and $2$. (bottom) Square magnetic
moment $\langle S^2 \rangle$ per site as function of the magnetic
field $h$ at $J=-1$. } \label{Fig.4}
\end{figure}

\begin{figure}[tbp]
\includegraphics[width=7cm]{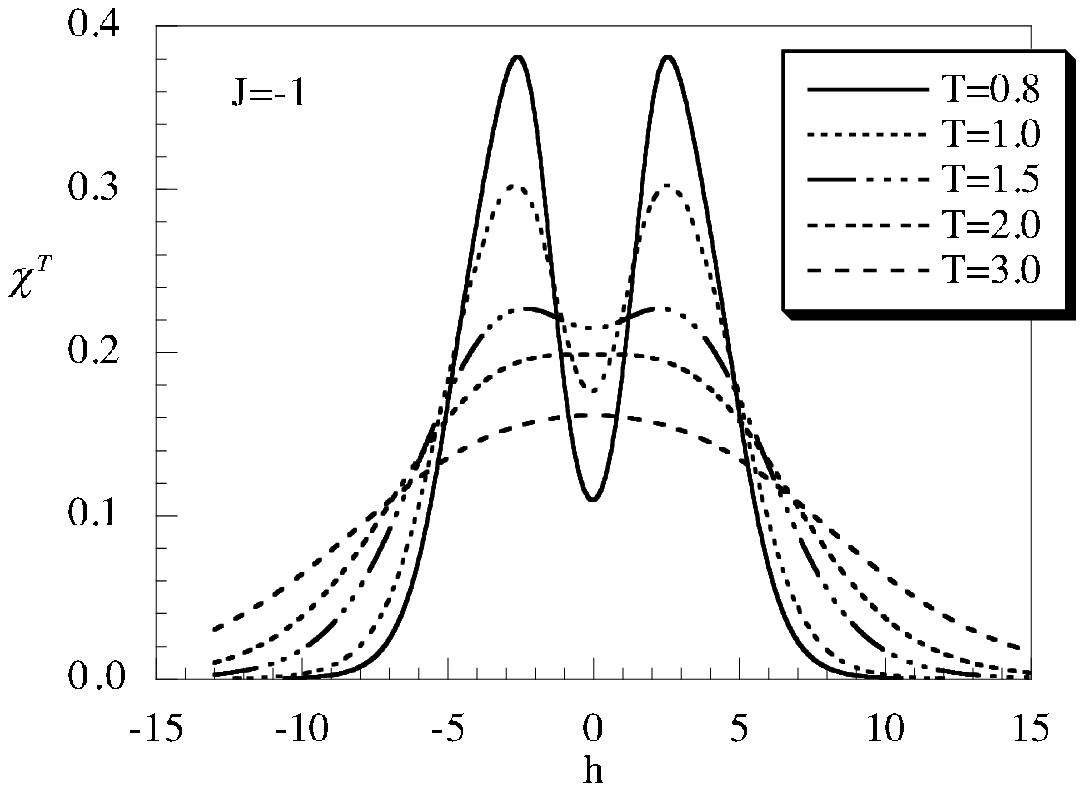}
\includegraphics[width=7cm]{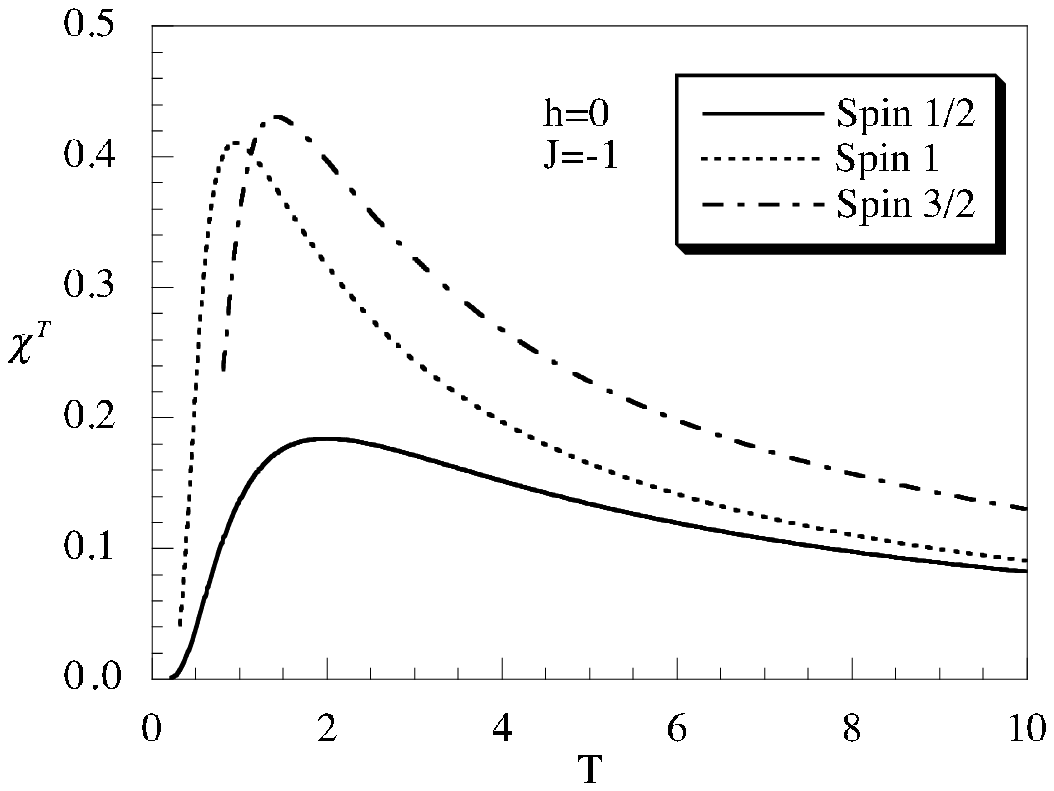}
\caption{Magnetic susceptibility $\chi$ at $J=-1$ as a function of
(top) external magnetic field $h$ for $T=0.8$, $1$, $1.5$, $2$, $3$
and (bottom) temperature $T$ for $h=0$ and spin $1/2$, $1$ and
$3/2$.} \label{Fig.5}
\end{figure}

\begin{figure}[tbp]
\includegraphics[width=7cm]{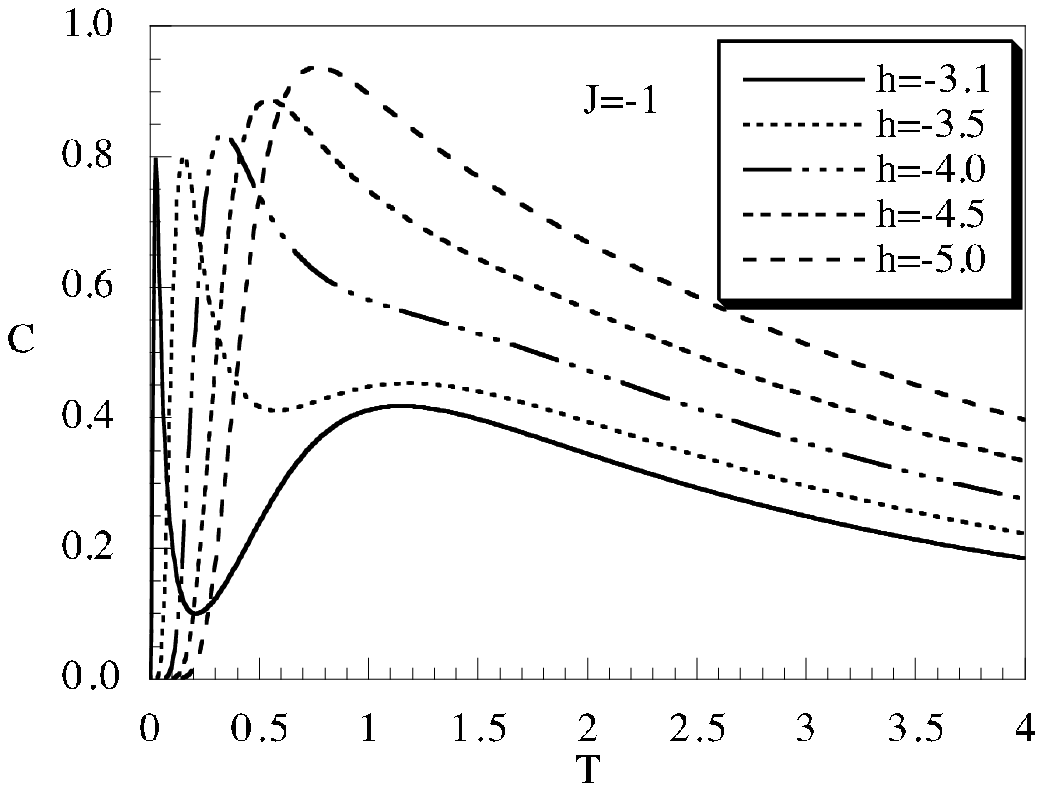}
\includegraphics[width=7cm]{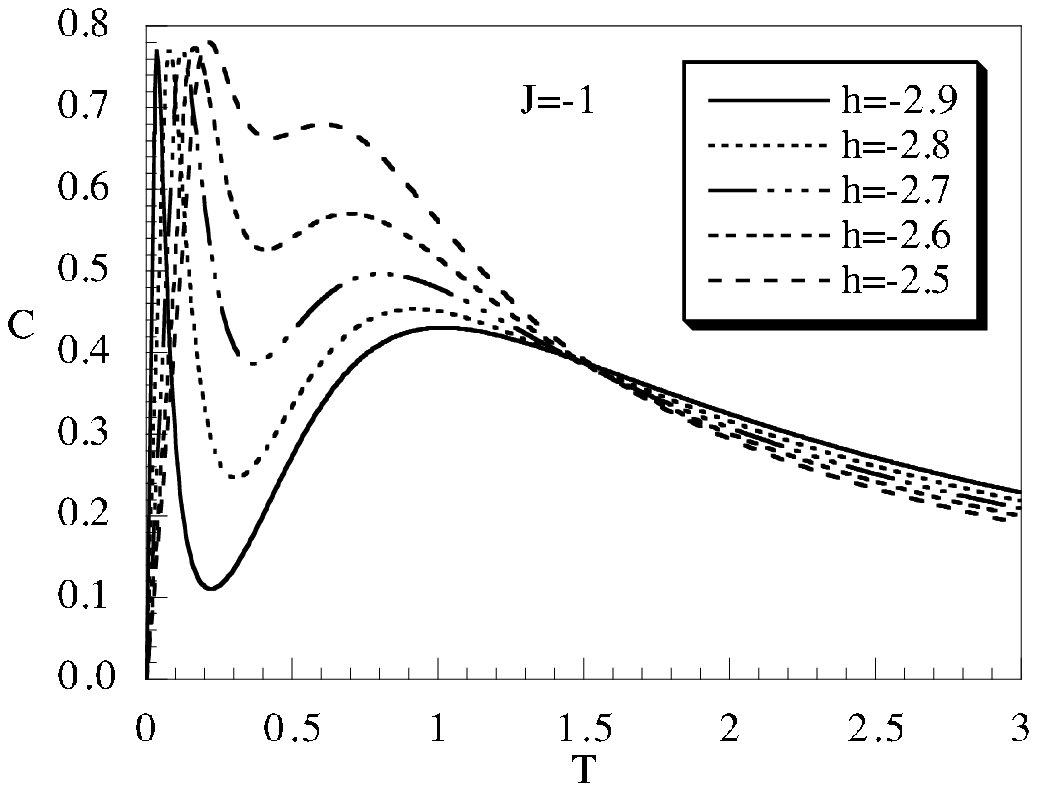}
\includegraphics[width=7cm]{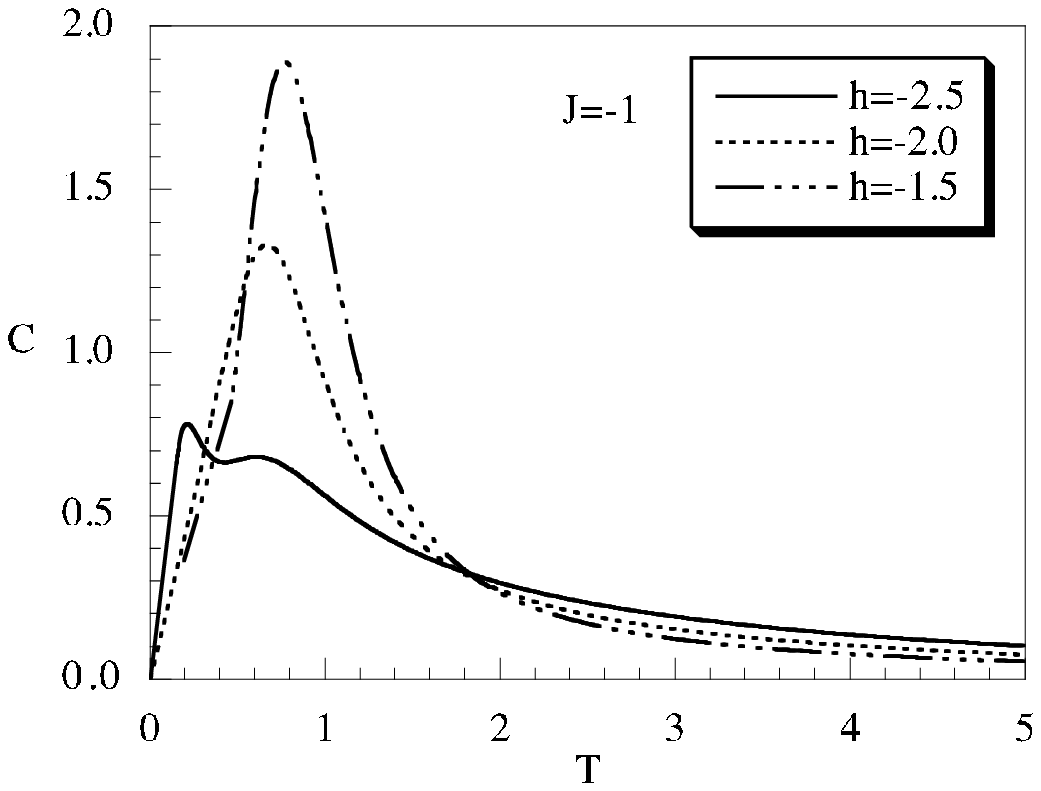}
\caption{Specific heat $C$ as a function of the temperature $T$ at
$J=-1$ and $h=-5\Rightarrow-1.5$.} \label{Fig.6}
\end{figure}

\subsection{Antiferromagnetic coupling $J=-1$}

In Fig.~\ref{Fig.4}, we report the magnetization per site $m$ as a
function of the magnetic field $h$ at $J=-1$, the square magnetic
moment $\langle S^2 \rangle$ per site as function of the magnetic
field $h$ at $J=-1$ and the double $D$ and the triple $T$
occupancies per site as functions of the chemical potential $\mu$ at
$V=1$ and $T=0$, $1$ and $2$. At zero temperature, going from
negative to positive values of the magnetic field, the magnetization
first jumps from $-\frac32$ to $0$ at $h=3J$ and then from $0$ to
$\frac32$ at $h=-3J$. At these values of the external magnetic
field, the state of lower energy changes from the fully polarized
one with spin $-\frac32$ ($E=0$) to the antiferromagnetically
aligned one with spins $\frac32$ and $-\frac32$
($E=-\frac32h+\frac92J$) and from this latter to the the fully
polarized one with spin $\frac32$ ($E=-3h$), respectively. The
double and the triple occupancies behave accordingly. The
temperature just makes the transition smoother as it results clear
by looking at the values of the square magnetic moment. These latter
also show the typical antiferromagnetic behavior that will be easily
identified in the features of the susceptibility discussed below.

In Fig.~\ref{Fig.5}, we report the magnetic susceptibility $\chi$ at
$J=-1$ as a function of the external magnetic field $h$ for $T=0.8$,
$1$, $1.5$, $2$, $3$ and the temperature $T$ for $h=0$ and spin
$1/2$, $1$ and $3/2$. As we could expect from the magnetization
curves, the susceptibility diverges, at zero temperature, at the
critical values discussed above and is null for all other values. On
increasing the temperature, the susceptibility becomes finite for
any value of the external magnetic field, although not negligible
only in an increasing, but finite, range of values. According to
this, as the integral of the curves over the whole axis should give
a constant value, $3$, that is, the larger jump in the
magnetization, the heights of the two peaks steadily decrease on
increasing the temperature. As a function of the temperature, the
susceptibility has the typical thermal activated behavior of the
parallel antiferromagnetic susceptibility for any value of the spin.
According to this, we can interpret the position of the maximum as
the temperature at which the system loses any memory of its
antiferromagnetic ground state at zero temperature.

In Fig.~\ref{Fig.6}, we report the specific heat $C$ as a function
of the temperature $T$ at $J=-1$ and $h=-5\Rightarrow-1.5$. The
curves seem to show the typical pattern caused by the interplay
between at least two gaps in the excitation spectrum: $h-3J$ and
$h-2J$. These would correspond to transitions between the fully
polarized state with spin $-\frac32$ and the antiferromagnetically
aligned state with spins $\frac32$ and $-\frac32$ and the fully
polarized state with spin $-\frac12$, respectively. The vanishing
value of the first gap, in the range of values of the external
magnetic field presented in the figures, could explain the much
larger sensitivity of the magnetization to temperature with respect
to what happens in the ferromagnetic state. It is worth noticing the
appearance of crossing points \cite{Vollhardt_97} for $3J<h<2.5J$
and $2.5J<h<1.5J$.

\section{Conclusions}

We have studied the Ising model with general spin $S$ in presence of
an external magnetic field by means of the equations of motion
method and of the Green's function formalism. First, the model has
been shown to be isomorphic to a fermionic one constituted of $2S$
species of localized particles interacting via an intersite Coulomb
interaction. Then, an exact solution has been found, for any
dimension, in terms of a finite, complete set of eigenoperators of
the latter Hamiltonian and of the corresponding eigenenergies. This
explicit knowledge has made possible writing exact expressions for
the corresponding Green's function and correlation functions, which
turn out to depend on a finite set of parameters to be
self-consistently determined. Finally, we have presented an original
procedure, based on algebraic constraints, to exactly fix these
latter parameters in the case of dimension $1$ and spin $\frac32$.
For this latter case and, just for comparison, for the cases of
dimension $1$ and spin $\frac12$ and spin $1$, relevant properties
such as magnetization $\langle S \rangle$ and square magnetic
moments $\langle S^2 \rangle$, susceptibility and specific heat are
reported as functions of temperature and external magnetic field
both for ferromagnetic and antiferromagnetic couplings. Ground state
properties and relevant transitions and gaps have been studied.
Crossing points in the specific heat have been identified.

\begin{widetext}

\appendix

\section{Algebraic relations}

\subsection{Formula for $n^{p}(i)$}

We want to find a recurrence relation for the operator $n^{p}(i)$,
with $ p\ge 1$. At first we note that
\begin{equation}
\begin{array}{l}
{n^{p}(i)=(n_{1}+n_{2}+n_{3})^{p}=\sum_{n=0}^{p}}\left(
\begin{array}{l}
p \\
n
\end{array}
\right) {\sum_{l=0}^{n}\left(
\begin{array}{l}
n \\
l
\end{array}
\right) n_{1}^{m-n}n_{2}^{n-l}n_{3}^{l}=[n_{1}^{p}+n_{2}^{p}+n_{3}^{p}]} \\
{+\sum_{n=1}^{p-1}\left(
\begin{array}{l}
p \\
n
\end{array}
\right)
[n_{1}^{p-n}n_{2}^{n}+n_{1}^{p-n}n_{3}^{n}+n_{2}^{p-n}n_{3}^{n}]+\sum
_{n=2}^{p-1}\left(
\begin{array}{l}
p \\
n
\end{array}
\right) \sum_{l=1}^{n-1}\left(
\begin{array}{l}
n \\
l
\end{array}
\right) n_{1}^{p-n}n_{2}^{n-l}n_{3}^{l}}
\end{array}
\label{A.1}
\end{equation}
Because of the property $n_{k}^{p}=n_{k}$, (\ref{A.1}) takes the form
\begin{equation}
n^{p}(i)=n(i)+D(i)\sum_{n=1}^{p-1}{\left(
\begin{array}{l}
p \\
n
\end{array}
\right) }+T(i)\sum_{n=2}^{p-1}{\left(
\begin{array}{l}
p \\
n
\end{array}
\right) }\sum_{l=1}^{n-1}{\left(
\begin{array}{l}
n \\
l
\end{array}
\right) }  \label{A.2}
\end{equation}
where $D(i)$ and $T(i)$ are the double and triple occupancy
operators as defined in (\ref{5.2}).

The sums in (\ref{A.2}) can be analytically performed
\begin{equation}
\begin{array}{l}
{b_{1}^{(p)}=\sum_{n=1}^{p-1}\left(
\begin{array}{l}
p \\
n
\end{array}
\right) =2^{p}-2} \\
{b_{2}^{(p)}=\sum_{n=2}^{p-1}\left(
\begin{array}{l}
p \\
n
\end{array}
\right) \sum_{l=1}^{n-1}\left(
\begin{array}{l}
n \\
l
\end{array}
\right) =3(1-2^{p}+3^{p-1})}
\end{array}
\label{A.4}
\end{equation}
and we have the algebraic relation
\begin{equation}
n^{p}(i)=n(i)+b_{1}^{(p)}D(i)+b_{2}^{(p)}T(i)  \label{A.5}
\end{equation}

\subsection{Formula for $[n^{\alpha }(i)]^{p}$}

By recalling that
\begin{equation}
[ n^{\alpha }(i)]={\frac{1}{2}}[n(i+a)+n(i-a)]  \label{A.6}
\end{equation}
we have for $p=2,3\cdots \cdots $
\begin{equation}
\begin{array}{l}
{[ n^{\alpha }(i)]^{p}={\frac{1}{{2^{p}}}}\sum_{m=0}^{p} \left(
\begin{array}{l}
p \\
m
\end{array}
\right) n^{p-m}(i+a)n^{m}(i-a)} \\
{={\frac{1}{{2^{p}}}}n^{p}(i+a)+{\frac{1}{{2^{p}}}}n^{p}(i-a)+{\frac{1}{{
2^{p}}}}\sum_{m=1}^{p-1}\left(
\begin{array}{l}
p \\
m
\end{array}
\right) n^{p-m}(i+a)n^{m}(i-a)}
\end{array}
\label{A.7}
\end{equation}
Because of the algebraic relation (\ref{A.5}) we obtain
\begin{equation}
[ n^{\alpha }(i)]^{p}={\frac{1}{2^{p}}}\sum
_{m=0}^{5}b_{m}^{(p)}Z_{m}(i)  \label{A.8}
\end{equation}
where the operators $Z_{m}(i)$ are defined as
\begin{equation}
\begin{array}{l}
{Z_{0}(i)=2n^{\alpha }(i)} \\
{Z_{1}(i)=2D^{\alpha }(i)+n(i+a)n(i-a)} \\
{Z_{2}(i)=2T^{\alpha }(i)+n(i+a)D(i-a)+n(i-a)D(i+a)} \\
{Z_{3}(i)=n(i+a)T(i-a)+D(i+a)D(i-a)+n(i-a)T(i+a)} \\
{Z_{4}(i)=D(i+a)T(i-a)+D(i-a)T(i+a)} \\
{Z_{5}(i)=T(i+a)T(i-a)}
\end{array}
\label{A.9}
\end{equation}
and the new coefficients $b_{m}^{(p)}$ have the expressions
\begin{equation}
\begin{array}{l}
{b_{0}^{(p)}=1} \\
{b_{3}^{(p)}=\sum_{m=1}^{p-1}\left(
\begin{array}{l}
p \\
m
\end{array}
\right) b_{2}^{(m)}=4(-1+2^{2p-2}+3\cdot 2^{p-1}-3^{p})} \\
{b_{4}^{(p)}=\sum_{m=1}^{p-1}\left(
\begin{array}{l}
p \\
m
\end{array}
\right) b_{1}^{(p-m)}b_{2}^{(m)}=5(1-2^{2p}-2^{p+1}+2\cdot 3^{p}-5^{p-1})}
\\
{b_{5}^{(p)}=\sum_{m=1}^{p-1}\left(
\begin{array}{l}
p \\
m
\end{array}
\right) b_{2}^{(p-m)}b_{2}^{(m)}=6(-1+5\cdot 2^{p-1}+5\cdot 2^{2p-1}-10\cdot
3^{p-1}-5^{p}+6^{p-1})}
\end{array}
\label{A.10}
\end{equation}
By solving (\ref{A.8}) with respect to the quantities $Z_{m}(i)$ we obtain
the recurrence relation
\begin{equation}
[n^{\alpha }(i)]^{p}=\sum_{m=1}^{6}A_{m}^{(p)}[n^{\alpha }(i)]^{m}
\label{A.11}
\end{equation}
where the coefficients $A_{m}^{(p)}$ are rational numbers defined as
\begin{equation}
\begin{array}{l}
{A_{1}^{(p)}={\frac{1}{{2^{p}}}}[2b_{0}^{(p)}-b_{1}^{(p)}+{\frac{2}{3}}
b_{2}^{(p)}-{\frac{1}{2}}b_{3}^{(p)}+{\frac{2}{5}}b_{4}^{(p)}-{\frac{1}{3}}
b_{5}^{(p)}]} \\
{A_{2}^{(p)}={\frac{1}{{2^{p}}}}[2b_{1}^{(p)}-2b_{2}^{(p)}+{\frac{{11}}{6}}
b_{3}^{(p)}-{\frac{5}{3}}b_{4}^{(p)}+{\frac{{137}}{{90}}}b_{5}^{(p)}]} \\
{A_{3}^{(p)}={\frac{1}{{2^{p}}}}[{\frac{4}{3}}b_{2}^{(p)}-2b_{3}^{(p)}+{
\frac{7}{3}}b_{4}^{(p)}-{\frac{5}{2}}b_{5}^{(p)}]} \\
{A_{4}^{(p)}={\frac{1}{{2^{p}}}}[{\frac{2}{3}}b_{3}^{(p)}-{\frac{4}{3}}
b_{4}^{(p)}+{\frac{{17}}{9}}b_{5}^{(p)}]} \\
{A_{5}^{(p)}={\frac{1}{{2^{p}}}}[{\frac{4}{{15}}}b_{4}^{(p)}-{\frac{2}{3}}
b_{5}^{(p)}]} \\
{A_{6}^{(p)}={\frac{1}{{2^{p}}}}{\frac{4}{{45}}}b_{5}^{(p)}}
\end{array}
\label{A.12}
\end{equation}
We note that
\begin{equation}
\begin{array}{l}
\sum_{m=1}^{6}A_{m}^{(p)}=1 \\
A_{m}^{(p)}=\delta _{mp}\quad (p\le 6)
\end{array}
\label{A.13}
\end{equation}
In table 1 we give some values of the coefficients
$A_{m}^{(p)}(p>6)$.

\begin{tabular}{|c|c|c|c|c|c|c|}
\hline $p$ & $A_{1}^{(p)}$ & $A_{2}^{(p)}$ & $A_{3}^{(p)}$ &
$A_{4}^{(p)}$ & $ A_{5}^{(p)}$ & $A_{5}^{(p)}$ \\ \hline $7$ &
$-\frac{45}{4}$ & $\frac{441}{8}$ & $-\frac{203}{2}$ &
$\frac{735}{8}$ & $-\frac{175}{4}$ & $\frac{21}{2}$ \\ \hline $8$ &
$-\frac{945}{8}$ & $\frac{9081}{16}$ & $-\frac{8085}{8}$ &
$\frac{13811 }{16}$ & $-\frac{735}{2}$ & $\frac{133}{2}$ \\ \hline
$9$ & $-\frac{5985}{8}$ & $\frac{56763}{16}$ & $-\frac{98915}{16}$ &
$\frac{ 81585}{16}$ & $-\frac{32739}{16}$ & $\frac{1323}{4}$ \\
\hline $10$ & $-\frac{59535}{16}$ & $\frac{559503}{32}$ &
$-\frac{480375}{16}$ & $ \frac{774575}{32}$ & $-\frac{37485}{4}$ &
$\frac{22827}{16}$ \\ \hline
\end{tabular}

\section{Algebra of the projection operators}

On the basis of the canonical anti-commutation relations (\ref{2.2})
it is straightforward to derive the algebra satisfied by the
operators $\xi_{a}(i)$, $\eta _{a}(i)$, $\zeta _{a}(i)$ defined in
(\ref{4.3}). Their anti-commutation relations are
\begin{equation}
\{\xi(i) ,\xi ^{\dagger }(j)\}= \delta_\mathbf{ij}\left(
\begin{array}{ccc}
{(1-n_{2}(i))(1-n_{3}(i))} & {-c_{1}(i)c_{2}^{\dagger
}(i)(1-n_{3}(i))} & {{{
-c_{1}(i)c_{3}^{\dagger }(i)(1-n_{2}(i))}}} \\
{-c_{2}(i)c_{1}^{\dagger }(i)(1-n_{3}(i))} &
{(1-n_{1}(i))(1-n_{3}(i))} & {{{
-c_{2}(i)c_{3}^{\dagger }(i)(1-n_{1}(i))}}} \\
{-c_{3}(i)c_{1}^{\dagger }(i)(1-n_{2}(i))} &
{{{-c_{3}(i)c{{_{2}^{\dagger }(i)}}(1-n_{1}(i))}}} &
{(1-n_{1}(i))(1-n_{2}(i))}
\end{array}
\right)  \label{B.3}
\end{equation}
\begin{equation}
\{\eta (i),\eta ^{\dagger (j)}\}=\delta_\mathbf{ij}\left(
\begin{array}{ccc}
{n_{2}(i)+n_{3}(i)-2n_{2}(i)n_{3}(i)} & {c_{1}(i)c_{2}^{\dagger
}(i)(1-2n_{3}(i))} & {{{
c_{1}(i)c_{3}^{\dagger }(i)(1-2n_{2}(i))}}} \\
{c_{2}(i)c_{1}^{\dagger }(i)(1-2n_{3}(i))} &
{n_{1}(i)+n_{3}(i)-2n_{1}(i)n_{3}(i)} & {{{
c_{2}(i)c_{3}^{\dagger }(i)(1-2n_{1}(i))}}} \\
{c_{3}(i)c_{1}^{\dagger }(i)(1-2n_{2}(i))} &
{{{c_{3}(i)c_{2}^{\dagger }(i)(1-2n_{1}(i))}}} & {
n_{1}(i)+n_{2}(i)-2n_{1}(i)n_{2}(i)}
\end{array}
\right)  \label{B.4}
\end{equation}
\begin{equation}
\{\zeta (i),\zeta ^{\dagger }(j)\}=\delta_\mathbf{ij}\left(
\begin{array}{ccc}
{n_{2}(i)n_{3}(i)} & {c_{1}(i)c_{2}^{\dagger }(i)n_{3}(i)} &
{{{c_{1}(i)c_{3}^{\dagger }(i)n_{2}(i)
}}} \\
{c_{2}(i)c_{1}^{\dagger }(i)n_{3}(i)} & {n_{3}(i)n_{1}(i)} &
{{{c_{2}(i)c_{3}^{\dagger }(i)n_{1}(i)
}}} \\
{c_{3}(i)c_{1}^{\dagger }(i)n_{2}(i)} & {c_{3}(i)c_{2}^{\dagger
}(i)n_{1}(i)} & {n_{1}(i)n_{2}(i) }
\end{array}
\right)  \label{B.5}
\end{equation}
It easy to verify that
\begin{equation}
\{\xi(i) ,\xi ^{\dagger }(j)\}+\{\eta(i) ,\eta ^{\dagger
}(j)\}+\{\zeta (i),\zeta ^{\dagger }(j)\}=\delta_\mathbf{ij}\left(
\begin{array}{ccc}
1 & 0 & 0 \\
0 & 1 & 0 \\
0 & 0 & 1
\end{array}
\right)  \label{B.6}
\end{equation}
Other relevant algebraic properties are
\begin{equation}
\begin{array}{l}
{{\sum_{a=1}^{3}\xi _{a}(i)\xi _{a}^{\dagger }(i)=3(1-n(i)+D(i)-T(i))}} \\
{{\sum_{a=1}^{3}\xi _{a}^{\dagger }(i)\xi
_{a}(i)=n(i)-2D(i)+3T(i)}\hfill }
\end{array}
\end{equation}
\begin{equation}
\begin{array}{l}
{{\sum_{a=1}^{3}\eta _{a}(i)\eta _{a}^{\dagger }(i)=2n(i)-4D(i)+6T(i)}} \\
{{\sum_{a=1}^{3}\eta _{a}^{\dagger }(i)\eta _{a}(i)=2D(i)-6T(i)}}
\end{array}
\end{equation}
\begin{equation}
\begin{array}{l}
{{\sum_{a=1}^{3}\zeta _{a}(i)\zeta _{a}^{\dagger }(i)=D(i)-3T(i)}} \\
{{\sum_{a=1}^{3}\zeta _{a}^{\dagger }(i)\zeta _{a}(i)=3T(i)}}
\end{array}
\label{B.7}
\end{equation}
By means of these relations we can express the particle density operator,
the double and triple occupancy operators as
\begin{equation}
\begin{array}{l}
{n(i)=3-\sum_{a=1}^{3}(\xi _{a}(i)\xi _{a}^{\dagger }(i)+\eta
_{a}(i)\eta
_{a}^{\dagger }(i)+\zeta _{a}(i)\zeta _{a}^{\dagger }(i))} \\
{D(i)=3-\sum_{a=1}^{3}(\xi _{a}(i)\xi _{a}^{\dagger
}(i)+{\frac{3}{2}}\eta
_{a}(i)\eta _{a}^{\dagger }(i)+2\zeta _{a}(i)\zeta _{a}^{\dagger }(i))} \\
{T(i)=1-\sum_{a=1}^{3}({\frac{1}{3}}\xi _{a}(i)\xi _{a}^{\dagger
}(i)+{\frac{1}{ 2}}\eta _{a}(i)\eta _{a}^{\dagger }(i)+\zeta
_{a}(i)\zeta _{a}^{\dagger }(i))}
\end{array}
\label{B.8}
\end{equation}
Also, the following relations hold
\begin{equation}
\begin{array}{l}
{{n^{2}(i)=n(i)+2D(i)}} \\
{{n(i)D(i)=2D(i)+3T(i)}\hfill } \\
{n(i)T(i)=3T(i)}
\end{array}
\qquad
\begin{array}{l}
{{D^{2}(i)=D(i)+6T(i)}\hfill } \\
{D(i)T(i)=3T(i)} \\
{{T^{2}(i)=T(i)}}
\end{array}
\label{5.19}
\end{equation}

\section{Calculation of $e^{-\beta H_{Ii}}$}

 From the definition $H_{Ii}=2Vn(i)n^{\alpha }(i)$we have
\begin{equation}
e^{-\beta H_{Ii}}=1+\sum_{p=1}^{\infty }{\frac{1}{{p!}}}
(-1)^{p}(2\beta V)^{p}n^{p}(i)[n^{\alpha }(i)]^{p}  \label{C.1}
\end{equation}
Recalling (\ref{A.5}) and (\ref{A.11}) we can write
\begin{equation}\label{C.3}
\begin{array}{l}
{e^{-\beta H_{Ii}}=1+\sum_{p=1}^{\infty }{\frac{1}{{p!}}}
(-1)^{p}(2\beta V)^{p}[n(i)+b_{1}^{(p)}D(i)+b_{2}^{(p)}T(i)]\sum
_{m=1}^{6}A_{m}^{(p)}[n^{\alpha }(i)]^{m}} \\
{=1+n(i)\sum_{m=1}^{6}f_{m}[n^{\alpha
}(i)]^{m}+D(i)\sum_{m=1}^{6}g_{m}[n^{\alpha
}(i)]^{m}+T(i)\sum_{m=1}^{6}h_{m}[n^{\alpha }(i)]^{m}}
\end{array}
\end{equation}
where
\begin{equation}\label{C.4}
\begin{array}{l}
{f_{m}=\sum_{p=1}^{\infty }{\frac{1}{{p!}}}(-1)^{p}(2\beta
V)^{p}A_{m}^{(p)}} \\
{g_{m}=\sum_{p=1}^{\infty }{\frac{1}{{p!}}}(-1)^{p}(2\beta
V)^{p}b_{1}^{(p)}A_{m}^{(p)}} \\
{h_{m}=\sum_{p=1}^{\infty }{\frac{1}{{p!}}}(-1)^{p}(2\beta
V)^{p}b_{2}^{(p)}A_{m}^{(p)}}
\end{array}
\end{equation}

By using the explicit expression (\ref{A.12}) of the coefficients
$A_{m}^{(p)}$, the infinite sums in (\ref{C.4}) can be analytically
performed. Straightforward calculations show that the parameters
$f_{m}$, $g_{m}$ and $h_{m}$ are linear combinations of $e^{-n\beta
V}$ with $n$ ranging from $0$ to $18$.

\end{widetext}


\end{document}